\documentclass[bibyear]{aa}  

\usepackage{graphicx}
\usepackage{natbib,twoopt}
\setcitestyle{aysep={}}
\usepackage{txfonts}
\usepackage{color}
\usepackage{tikz}
\def\ckmk{\tikz\fill[scale=0.4](0,.35) -- (.25,0) -- (1,.7) -- (.25,.15) -- cycle;}

\usepackage[colorlinks=true,citecolor=blue,linkcolor=blue,urlcolor=blue]{hyperref}
\usepackage{lscape}

\begin{document}

   \title{The Belgian repository of fundamental atomic data and stellar spectra (BRASS)}
        \subtitle{II. Quality assessment of atomic data for unblended lines in FGK stars\thanks{Tables \ref{loggfperstar}, \ref{loggfperstar2}, \ref{equivalentwidthexample}, and \ref{appendixtable} are available in full, in electronic format via the CDS and at \href{http://brass.sdf.org}{brass.sdf.org}.}	}

   \author{M. Laverick\inst{1,2},
          A. Lobel\inst{2},
          P. Royer\inst{1},
          T. Merle\inst{3},
          C. Martayan\inst{4},
          P.A.M. van Hoof\inst{2},
          M. Van der Swaelmen\inst{3},
          M. David\inst{5},
          H. Hensberge\inst{2}, and 
          E. Thienpont\inst{6}
          }

   \institute{Instituut voor Sterrenkunde, KU~Leuven, Celestijnenlaan 200D, box 2401, 3001 Leuven, Belgium
            \and
             Royal Observatory of Belgium, Ringlaan 3, B-1180 Brussels, Belgium \\
                \email{mike.laverick@kuleuven.be}
         \and
              Institut d'Astronomie et d'Astrophysique, Universit\'{e} Libre de Bruxelles, Av. Franklin Roosevelt 50, CP 226, 1050 Brussels,
         \and
             European Organisation for Astronomical Research in the Southern Hemisphere, Alonso de C\'{o}rdova 3107, Vitacura, 19001 Casilla, Santiago de Chile, Chile
 Belgium
         \and
             Onderzoeksgroep Toegepaste Wiskunde, Universiteit Antwerpen, Middelheimlaan 1, 2020 Antwerp, Belgium,
         \and Vereniging voor Sterrenkunde, Kapellebaan 56, 2811 Leest, Belgium
             }

   \date{Received June 4th, 2018; accepted February 5th, 2019}

% \abstract{}{}{}{}{} 
% 5 {} token are mandatory
 
  \abstract
   {Fundamental atomic transition parameters, such as oscillator strengths and rest wavelengths, play a key role in modelling and understanding the chemical composition of stars in the universe. Despite the significant work under way to produce these parameters for many astrophysically important ions, uncertainties in these parameters remain large and can limit the accuracy of chemical abundance determinations.} 
   {The Belgian repository of fundamental atomic data and stellar spectra (BRASS) aims to provide a large systematic and homogeneous quality assessment of the atomic data available for quantitative spectroscopy. BRASS shall compare synthetic spectra against extremely high-quality observed spectra, at a resolution of $\sim$85000 and signal-noise ratios of $\sim$1000, for approximately 20 bright BAFGK spectral-type stars, in order to critically evaluate the atomic data available for over a thousand potentially useful spectral lines.
}
   {A large-scale homogeneous selection of atomic lines is performed by synthesising theoretical spectra of literature atomic lines for FGK-type stars including the Sun, resulting in a selection of 1091 theoretically deep and unblended lines in the wavelength range 4200-6800~\AA, which may be suitable for quality assessment. Astrophysical $\log(gf)$ values are determined for the 1091 transitions using two commonly employed methods. The agreement of these $\log(gf)$ values are used to select well-behaved lines for quality assessment. }
   { We found 845 atomic lines to be suitable for quality assessment, of which 408 were found to be robust against systematic differences between analysis methods. Around {53\%} of the quality-assessed lines were found to have at least one literature $\log(gf)$ value in agreement with our derived values, though the remaining values can disagree by as much as 0.5~dex. Only $\sim$38\% of \ion{Fe}{I} lines were found to have sufficiently accurate $\log(gf)$ values, increasing to $\sim$70-75\% for the remaining Fe-group lines. }
   {}
   \authorrunning{Laverick et al}
   \titlerunning{BRASS II. Quality assessment of atomic data for unblended lines in FGK stars}

   \maketitle

\section{Introduction}
Modern quantitative stellar spectroscopy routinely measures spectral line properties, such as elemental abundances, to a precision of 0.01~dex. However, systematic errors often dominate derived spectral line properties resulting in actual errors that are typically an order of magnitude larger \citep{barklem2016}. The work of \cite{lindegren2013} shows that the uncertainties in elemental abundances are a limiting factor in discriminating stellar populations, and that small improvements in uncertainty can significantly reduce the number of stars required for galactic archaeology. Such improvements are important for current and upcoming large-scale galactic surveys such as Gaia \citep{gaia2016}, GALAH \citep{galah2015}, WEAVE \citep{weave2012}, and 4MOST \citep{4most2012}.

One of the main uncertainties in elemental abundances is the quality of the adopted atomic data used in spectral synthesis calculations \citep{bigot2008}.
 The Belgian repository of fundamental atomic data and stellar spectra (BRASS) aims to provide astronomers with quality information for the large amount of atomic data available for high-resolution optical spectroscopy, in an attempt to help reduce systematic input errors in quantitative spectroscopy from atomic data and line selection \citep{alexbrass}. Previously, we retrieved and cross-matched a large quantity of atomic data from several major atomic databases such as the Vienna Atomic Line Database (VALD3; \citealp{vald3}), the National Institute of Standards and Technology Atomic Spectra Database (NIST~ASD; \citealp{nist}), and providers within the Virtual Atomic and Molecular Data Centre (VAMDC; \citealp{vamdc}), in preparation for quality assessment work (\citealp{brass1}; hereafter Paper~1). In this work the atomic data of seemingly `unblended' spectral lines are quality assessed against several benchmark dwarf stars, including the Sun, spanning late F-type to early K-type stars, for the spectral range 4200-6800~\AA.\footnote{Throughout this paper we use the term `unblended' to refer to spectral lines that produce the majority of the flux absorption of their respective spectral feature, in the context of theoretical spectral calculations, as explained in Section 3.1.} Unblended spectral lines for stars of $\sim$G2V spectral type are identified in a homogeneous manner using both the observed benchmark spectra, and the theoretical input line list of Paper~1. Astrophysical oscillator strengths\footnote{often written in terms of log weighted-oscillator strengths, or $\log(gf).$} are derived for these unblended lines, using two commonly utilised methods, to gauge the reliability of the spectral line for quantitative spectroscopy and to produce `benchmark' $\log(gf)$ values for quality assessment of atomic data. The literature $\log(gf)$ values are then compared against these benchmark $\log(gf)$ values to determine which literature values reliably reproduce the stellar spectra of cool dwarf stars, and thus whether the values can be recommended for spectroscopic modelling. This paper presents three main sets of results:
\begin{itemize}
\item a set of extremely high-quality stellar spectra, with signal-to-noise ratios of S/N$\sim$800-1200 and a resolution R$\sim$85000, for six well-studied stars between 5000-6000~K;
\item an identification of unblended and reliable atomic lines in these stellar spectra, determined using theoretical input line lists and the behaviour of the observed spectra;
\item a set of benchmark $\log(gf)$ values for these unblended lines, including a quality assessment of the available atomic data $\log(gf)$ values per spectral line.
\end{itemize}

Section~2 outlines the selection of the 5000-6000~K benchmark stars, the data reduction of the observed spectra, the details of the spectral synthesis calculations, and the stellar parameter determination of the benchmark stars. Section~3 covers the selection of theoretically unblended spectral lines and the measurement of their observed counterparts. Section~4 details the determination of astrophysical oscillator strengths for the selected lines, as well as their uncertainties. Finally, Section~5 explains how the derived oscillator strengths and uncertainties are used as benchmark values to perform a quality assessment of the available literature atomic data for the selected spectral lines. The findings of our quality assessment for over 800 spectral lines are also discussed.
\section{Benchmark stars between 5000 and 6000~K}

\subsection{Selection and observations of benchmark stars}

\begin{table*}
\centering
\caption{Stellar parameters of the FGK benchmark stars with effective temperatures in the range 5000~$\leq T_{\mathrm{eff}} \leq$~$\sim$6000~K. The associated stellar parameter errors are  $T_{\mathrm{eff}}~\pm 50$~K, $\log{g}~\pm 0.20$~dex, [M/H]~$\pm$~0.01~dex, $\zeta_{\mu}~\pm$~0.10~kms$^{-1}$, and $v\sin{i}~\pm$~1.00~kms$^{-1}$.
}
\begin{tabular}{lccccc}
            \hline
            \hline
            \noalign{\smallskip}
            Object      &  $T_{\mathrm{eff}}$ (K) & $\log g$ (dex) & [M/H] (dex) & $\zeta_{\mu}$ (kms$^{-1}$) & $v\sin i$ (kms$^{-1}$)   \\
            
                        &$\pm50$ & $\pm0.20$ & $\pm0.01$ & $\pm0.10$ & $\pm1.00$ \\
            
            \noalign{\smallskip}
            \hline
            \noalign{\smallskip}
                        $\beta$ Com & 6010 & 4.35 &  0.06 & 1.14 & 5.70 \\
                    10 Tau & 5912 & 3.90 & -0.11 & 1.27 & 5.25 \\
                    51 Peg & 5804 & 4.42 &  0.20 & 1.10 & 4.00 \\           
            Sun & 5777 & 4.44 &  0.00 & 1.10 & 2.50 \\
                    70 Vir   & 5500 & 3.94 & -0.11 & 1.05 & 3.60 \\
            70 Oph A    & 5354 & 4.60 &  0.07 & 0.96 & 3.15 \\
                        $\epsilon$ Eri & 5136 & 4.71 & -0.03 & 0.90 & 3.40\\          
            \hline            
\end{tabular}
\label{astroparams}
\end{table*}

For the benchmark star selection the following criteria were imposed: The stars must
\begin{itemize}

\item fall within the temperature range 5000-6000~K;
\item not belong to any known stellar variability classes;
\item not be chemically peculiar or metal-poor, \textit{i.e.}  have roughly solar-like abundances; 
\item have low $v\sin i$-values so that spectral lines are narrow;
\item ideally be dwarf stars to avoid non-LTE effects due to extended atmospheres;
\item be brighter than seventh magnitude in $V$ so that an extremely high-S/N spectrum of the object can be obtained.
\end{itemize}

Using these criteria the following stars were chosen: $\epsilon$~Eri, 70~Oph~A, 70~Vir, 51~Peg, 10~Tau, $\beta$~Com, and the Sun. The benchmark spectra were obtained using the HERMES echelle spectrograph mounted on the Mercator 1.2~m telescope at the Roche de los Muchachos Observatory, La Palma, Spain, which is able to observe the complete wavelength range 3800-9000~\AA\ in a single exposure at a resolution of R$\sim$85000 \citep{hermes}.
Each object was observed 10-50 times, depending on the $V$~mag, in succession throughout a single night. Extra flat-fields were also taken during the night to avoid introducing systematic noise in the reduction process. The resultant exposures were reduced using the dedicated HERMES pipeline (release V6.0) and  co-added to produce a single stacked spectrum per object with a S/N of $\sim$800-1200. The solar spectrum was obtained from \cite{solarfts}, taken using the NSO/KPNO Fourier Transform Spectrograph (FTS) with a spectral resolution of R$\sim$350000 and a S/N of $\sim$2500. The wavelength range in this work is limited to 4200-6800~\AA\ to avoid heavily blended features in shorter wavelengths, and to avoid telluric contamination in the longer wavelengths.

\subsection{Theoretical spectral modelling}

The synthesis work is performed using the atomic line list detailed in Paper~1, and 1D hydrostatic, plane-parallel atmospheric models computed using the publicly available ATLAS9 code \citep{atlas9}. The models adopt the updated opacity distribution functions (ODFs) calculated by \cite{castelliopacity}, use the "mixing length" approximation detailed by \citet{castelli1997} in conjunction with a mixing length of $L/H_{\rho}=1.25$, and with the ATLAS9 "approximate treatment of overshooting" turned off as recommended by \citet{bonifacio2012}. The model grid of \cite{castelli1997}, calculated with the updated ODFs, was used as a starting point to calculate a much finer grid of stellar models in order to derive stellar parameters for the benchmark spectra. The grid stellar parameters cover 4000~$< T_{\mathrm{eff}} <$~15000~K (step size of 50~K), $0.0 < \log g < 5.0$~dex (step size of 0.1 dex),  $-5.0 < $~[M/H]~$ < 1.0$~dex (step size of 0.2 dex), 0~$ < \zeta_{\mu} <$~20~kms$^{-1}$ (step size of 0.5~kms$^{-1}$), and 0~$ < v\sin i <$~300~kms$^{-1}$ (step size of 1~kms$^{-1}$).

The calculated models used the same mixing length and overshooting treatment as Castelli, but adopt the solar abundances of \cite{solar2007} to be consistent with the transfer calculations. All synthetic spectra are computed in local thermal equilibrium (LTE), using the publicly available radiative transfer code TurboSpectrum V12.1.1 \citep{turbospectrum, plez2012}. The synthetic spectra are computed using the solar abundances of \cite{solar2007}. We assume all benchmark stars to have the same elemental abundance distribution as the Sun, but scaled according to the metallicity of the given benchmark star. The spectra are rotationally broadened to the $v\sin i$ of the given star, and instrumentally broadened to the spectral resolution of HERMES (or the FTS for the solar spectrum). Finally, the SpectRes Python resampling tool was used when comparing synthetic and observed spectra \citep{carnall2017}. The choice of synthesis code and stellar models was governed by the aim of the project, namely to constrain the atomic data of many lines, as homogeneously as possible, for stars of B-, A-, F-, G-, and K-spectral types. Section 4.3.2. provides a brief discussion of some of the uncertainties in the adopted modelling procedure.

\subsection{Stellar parameter determination}

For consistency the spectral parameters of the six benchmark stars observed by HERMES, initially outlined in \cite{alexbrass}, were determined as follows, taking account of the spectral modelling details discussed in Section~2.2. The stellar parameters are initially estimated using a limited number of diagnostic Balmer, Fe, and Mg absorption lines. The method then iterates over $T_{\mathrm{eff}}$, $\log{g}$, [M/H], and $\zeta_{\mu}$ until the best fit is found to the detailed shapes of a more extensive set of $\sim$30 diagnostic \ion{Fe}{I} and \ion{Fe}{II} photospheric lines. The fit method iterates until the best fit to the continuum-normalised Fe-line profiles has been accomplished using $\chi^2$ minimisation. The iterations over the $\zeta_{\mu}$ proceed until the iron abundance value determined from the \ion{Fe}{I} lines is in agreement with the iron abundance value determined from the \ion{Fe}{II} lines. The iteration procedure assumes the $\zeta_{\mu}$ stays constant with depth in the line formation regions of these medium-strong Fe lines. During each iteration, the spectrum is continuum-flux normalised using a semi-automatic template normalisation routine described in Appendix~A.

The $v\sin{i}$ value, that is, the convolution of the projected rotational velocity and macro-broadening, is also iterated in steps of 1~kms$^{-1}$ while obtaining the best fit to the detailed Fe line profiles. The final stellar parameter errors are:  $T_{\mathrm{eff}} \pm 50$~K, $\log{g} \pm 0.20$~dex, [M/H]~$\pm$~0.01~dex, $\zeta_{\mu} \pm$~0.10~kms$^{-1}$, and $v\sin{i} \pm$~1.00~kms$^{-1}$. The standard solar parameters of $T_{\mathrm{eff}} = 5777$~K and $\log{g} = 4.44$~dex were adopted, with the remaining parameters determined in this work. The FTS solar spectrum normalisation is described by \cite{solarfts}. The stellar parameters of the benchmark stars are listed in Table~\ref{astroparams}, and are found to be in good agreement with other literature surveys deriving astrophysical parameters for these stellar objects \citep{dasilva2015,brewer2016}.

\section{Selection of unblended spectral lines}

\subsection{Quantification of theoretical line blending}

\begin{figure*}
\centering
\includegraphics[width=18cm]{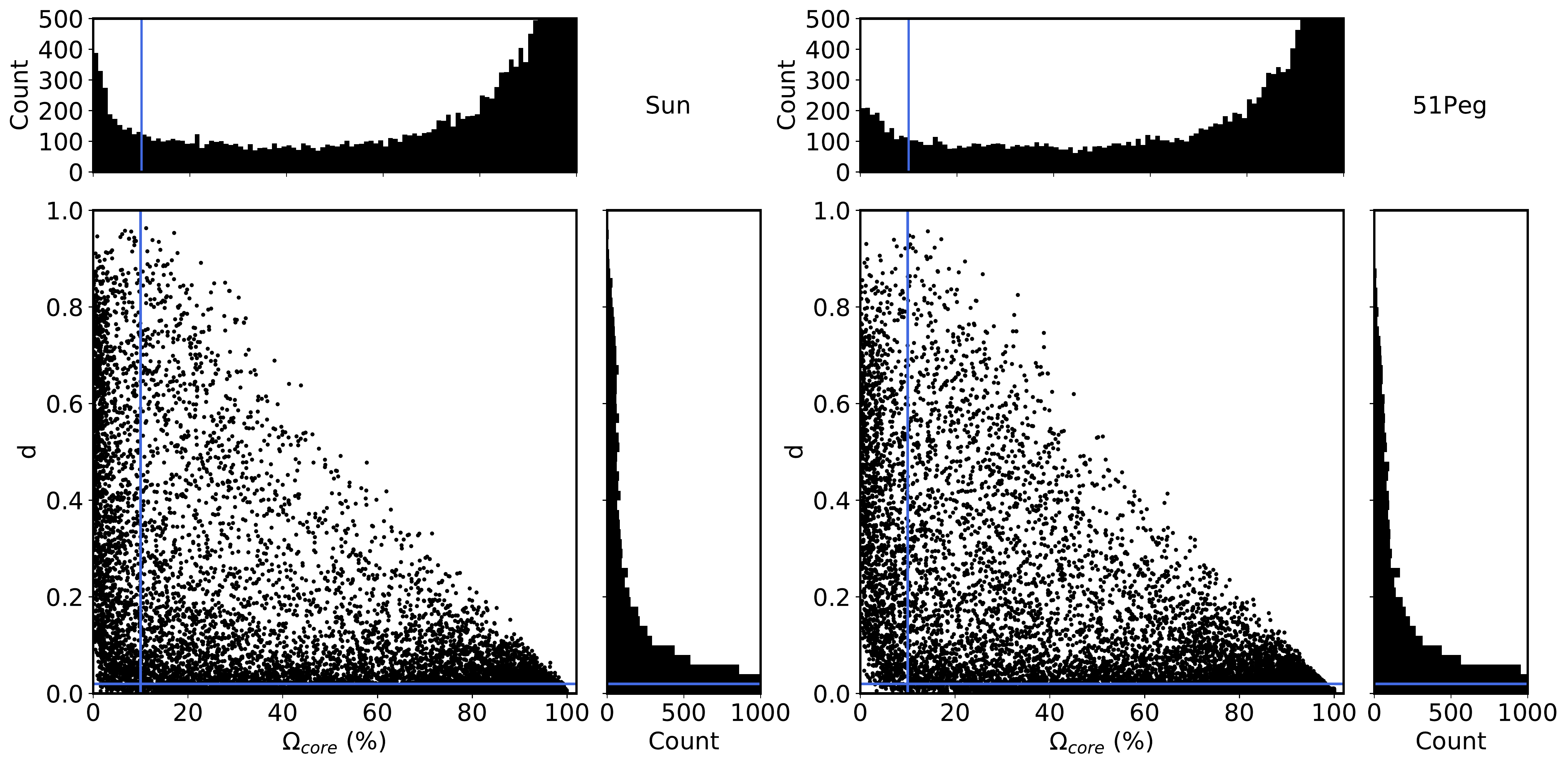}
\caption{Spectral line depth, $d$, plotted against theoretical core-blending, $\Omega_{core}$, (defined in Eqs.~1~and~2) for all 82337 lines of the input atomic line list for the Sun~(G2V) and 51~Peg~(G2V). Solid blue lines show the selected cut-offs for the 5000-6000~K benchmark stars of $d \geq 0.02$ and $\Omega_{core} \leq 10 \%$. Analysis of hotter benchmark stars will likely need different cut-offs, or potentially a different approach to treating blended spectral lines.}
\label{blenddepth}
\end{figure*}

We define the theoretical core blending, $\Omega_{core}$, of a given line as follows:

\begin{equation}
\begin{gathered}
      \Omega_{core} = 1 - \frac{\int^{\lambda_0+x}_{\lambda_0-x}~\left( 1 - F(\lambda)_{norm}^{~line}\right)~\mathrm{d\lambda}}{ \int^{\lambda_0+x}_{\lambda_0-x}~\left( 1 - F(\lambda)_{norm}^{~tot}~\right)~\mathrm{d}\lambda }  
\end{gathered}
,\end{equation}
where
\begin{equation}
\begin{gathered}
      x = \frac{1}{2} \frac{W_{\lambda}^{~line}}{d}
\end{gathered}
,\end{equation}

where $W_{\lambda}^{~line}$ is the equivalent width of the given line, $d$ is the central depth of the line relative to the normalised continuum, $\lambda_0$ is the rest wavelength of the line, $F(\lambda)_{norm}^{~line}$ is the normalised flux of the line, and $F(\lambda)_{norm}^{~tot}$ is the normalised flux of the total spectrum, including all other spectral features.  A line with $d~=~0$ does not appear in the spectrum, whereas a line with $d~=~1$ absorbs all available flux at $\lambda_0$. A line with $\Omega_{core} = 0$ (or 0\%) is considered completely unblended in its core wavelength region $\lambda_0 - x$ to $\lambda_0 + x$, whereas a line with  $\Omega_{core} \approx 1$  (or $\sim$100\%) is completely blended with other spectral features.

The core blending, $\Omega_{core}$, is calculated for all 82337 lines in the input atomic line list, for the spectra of the Sun and 51~Peg at the resolutions of R$\sim$350000 and R$\sim$85000 respectively. Figure~\ref{blenddepth} shows the $\Omega_{core}$ as a percentage, plotted against the normalised central line depth, $d$, for the Sun and 51~Peg. Both the Sun and 51~Peg exhibit the same distributions: a substantial number of shallow, heavily-blended \textit{background} lines, in addition to a number of relatively unblended lines. To reduce the impact of blending on the quality assessment (discussed later in Section~4.3.7) a cut-off of $\Omega_{core}$~$\leq 10\%$ is imposed, marked by the blue vertical lines in Fig.~\ref{blenddepth}. This cut-off was selected as a balance between $\Omega_{core}$ and the number of investigated lines, and therefore it encompasses the full peak in unblended lines for 51~Peg. An additional cut-off on central line depth $d \geq 0.02$ is imposed to help ensure that observed line profiles are actually measurable and less affected by any noise in the observed spectra.

Using these constraints, 1515 atomic lines are retained for 51~Peg, and 1954 lines in the Sun. All 1515 lines are present in both 51~Peg and the Sun, the difference in quantity being attributed to the differing resolutions of the two spectra. Given that the majority of the benchmark spectra are taken with HERMES, the 1515 spectral lines were selected for measurements, and are henceforth described as `unblended' lines. No limits have been placed on the investigated species, other than mandating that they must fall in the investigated wavelength range, and must be deep and relatively unblended according to the theoretical line list.

\subsection{Measurement of observed line profiles}

The 1515 unblended lines were automatically measured in each of the seven observed benchmark spectra using a single Gaussian profile fit to determine if the feature exists and to determine its equivalent width, $W_{\lambda}$.

\begin{figure}
\centering
\includegraphics[width=9.5cm]{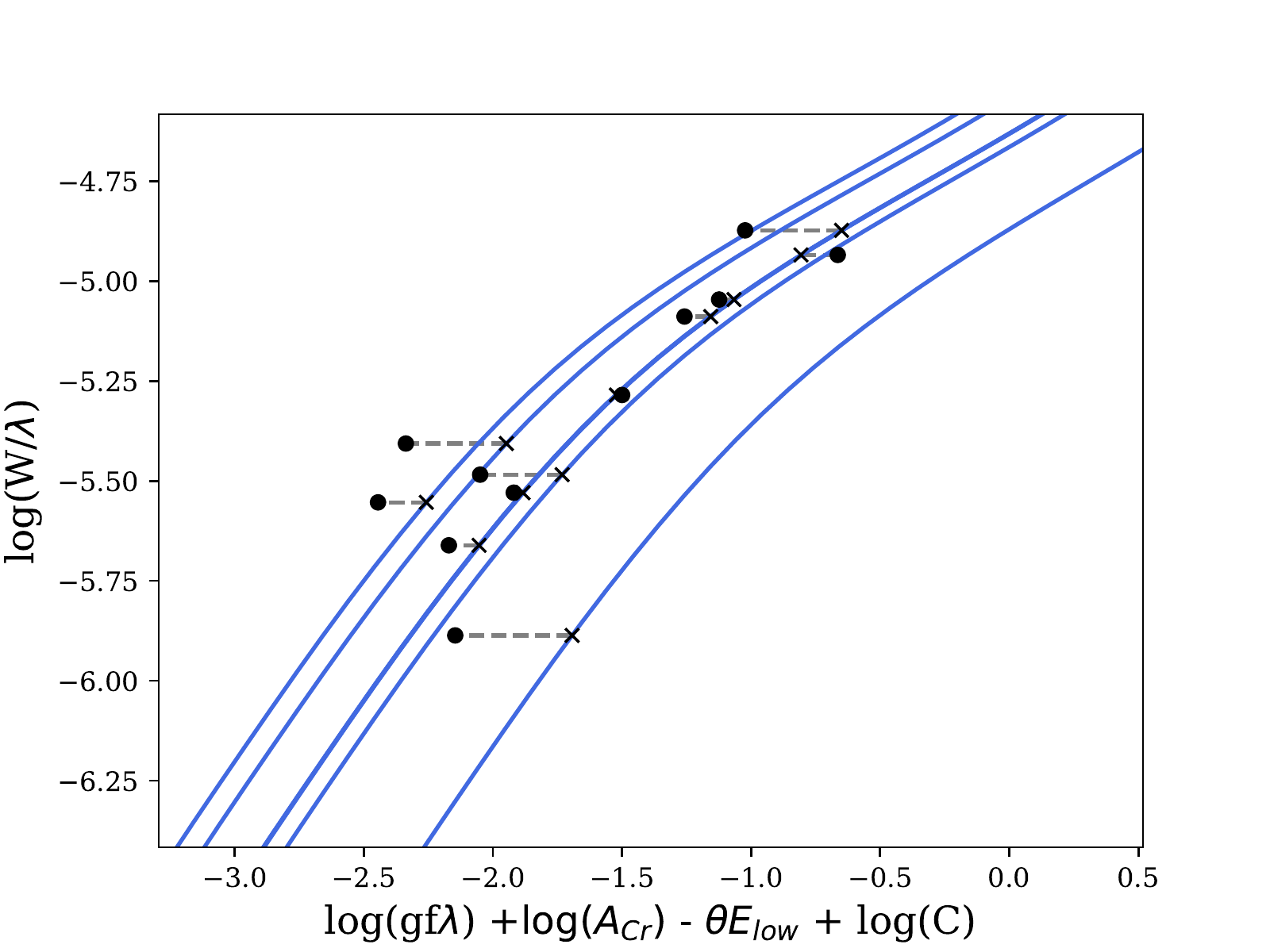}
\caption{Conversion of observed \ion{Cr}{II} line equivalent width values into $\log(gf)$ values using theoretical curves of growth for a given star. Blue lines represent theoretical curves of growth for different \ion{Cr}{II} lines. Black circles are measured equivalent width values plotted using the $\log(gf)_{input}$ values, while black crosses are measured equivalent width values plotted using derived $\log(gf)$ values, \textit{i.e.} the intersect of the equivalent width value with its respective curve of growth.}
\label{cog_method}
\end{figure}

The Gaussian fit is optimised using Gauss-Newton non-linear regression, or a Nelder-Mead minimisation in the case of slow convergence, as described by \cite{wao_alex}. The best fit to the observed line flux is limited to the wavelength interval between the two local flux maxima either side of $\lambda_0$ that exceed 2\% of the normalised continuum flux level. A goodness-of-fit value of $\chi^2 \geq$~0.95, in addition to manual inspection, was used to filter out poorly fitted, non-existent, heavily blended, or telluric-contaminated features. After measurements, 1091 spectral lines were found to be suitable for astrophysical $\log(gf)$ determination. A small number of these spectral lines cannot be assessed in all seven benchmark stars due to poor fitting in only some of the profiles, however the line can still be investigated at the cost of increased statistical uncertainty, discussed in Section~4.3.8. Additionally, the adoption of a Gaussian fit in place of a Voigt profile fit is discussed in Section~4.3.5.

\section{Determination of astrophysical oscillator strengths for selected lines}

There are two commonly employed methods to determine astrophysical $\log(gf)$ values: measuring a line equivalent width, $W_{\lambda}$, and converting it into a $\log(gf)$ value via the curve of growth \citep[\textit{e.g.}][]{sousa2007,onehag2012,Andreasen2016}, or by using detailed transfer calculations to vary a $\log(gf)$ value and determine a best-fitting value \citep[\textit{e.g.}][]{boeche2016,Tsantaki2018}. Both these methods are subject to a number of assumptions, especially in their treatment of blended lines, which can lead to significant systematic differences in derived $\log(gf)$ values if not properly accounted for. As such, both the curve of growth and iterative modelling methods have been explored, with the following section detailing their implementation and associated uncertainties.

\subsection{Determination of oscillator strengths using theoretical curves of growth}

The measured $W_{\lambda}$ value of a spectral line can be converted into a $\log(gf)$ value, on a star-by-star basis, using the following curve of growth relationship:
 
\begin{equation}
\begin{gathered}
      \log \left (\frac{W_{\lambda}}{\lambda_0} \right ) = \log(gf\lambda_0) + \log(A_{el})  - \frac{5040}{T_e}E_{low} +\log(C)
\end{gathered}
,\end{equation}
where $W_{\lambda}$ is the equivalent width, $\lambda_0$ is the transition wavelength, $A_{el}$ is the elemental abundance of the species, $T_e$ is the excitation temperature, $E_{low}$ is the transition lower energy level, and $C$ contains the remaining terms such as the Saha population factor and the continuum opacity. Theoretical curves of growth are built by varying the $\log(gf)$ value of each of the 1091 selected lines over the range -6.0~<~$\log(gf)$~<~4.0~dex in steps of 0.1~dex, and then calculating the equivalent widths of the resultant line profiles. The curves are synthesised using the rest-wavelength and energy levels of the selected line, for each of the seven benchmark stars, resulting in a total of 7658 individual curves of growth. A high-order polynomial is fitted through each of the curves, with maximum fit deviation of $\Delta\log(gf)$~$\ll$~0.005~dex. The measured $W_{\lambda}$ values are converted into $\log(gf)$ values using their respective curve of growth per line per star, as shown in Fig.~\ref{cog_method}. These individual $\log(gf)_{cog}$ values per star are averaged over the benchmark stars to produce a final mean $\overline{\log(gf)}_{cog}$ value and associated uncertainty, as discussed in Section~4.3.8.

\begin{figure*}
\centering
\includegraphics[width=18cm]{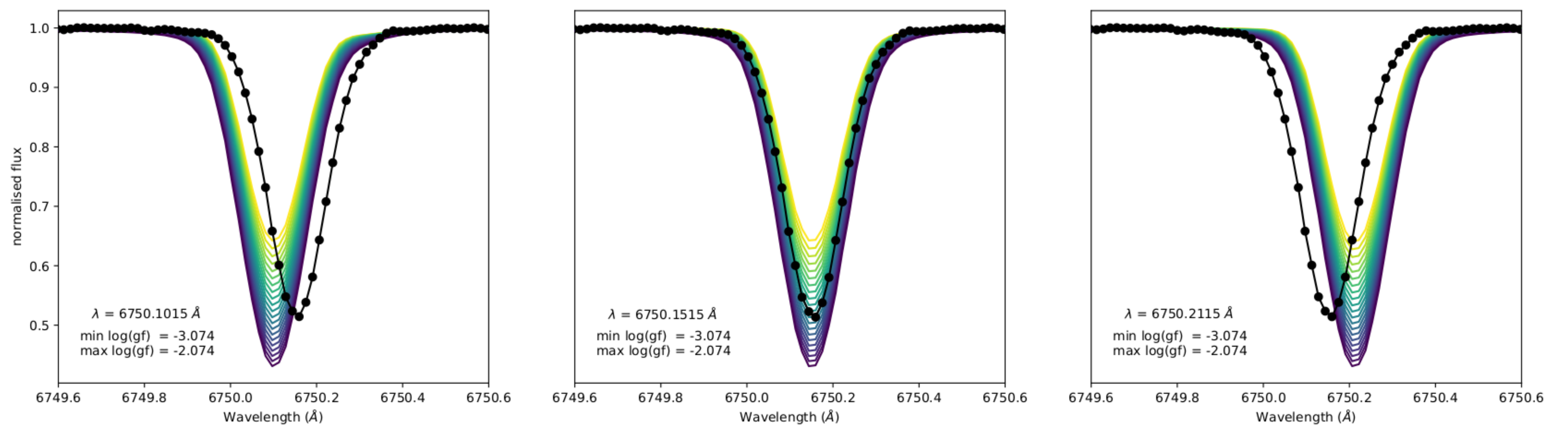}
\caption{An example of the iterative detailed modelling of the $\lambda6750.2$ \ion{Fe}{I} line in 51~Peg. The observed line profile is shown in black with solid markers, and the line profiles synthesised with varying input atomic parameters are shown in coloured lines (corresponding to different $\log(gf)$ values). The panels show three of the synthesised wavelength values: $\lambda_{input}$~-0.05~\AA\ (left), $\lambda_{input}$ (middle), and $\lambda_{input}$~+0.05~\AA\ (right). Each panel shows the full -0.5~dex~<~$\log(gf)$~<~+0.5~dex grid (in steps of 0.05~dex).  }
\label{gridprofiles}
\end{figure*}

\begin{figure*}
\centering
\includegraphics[width=18cm]{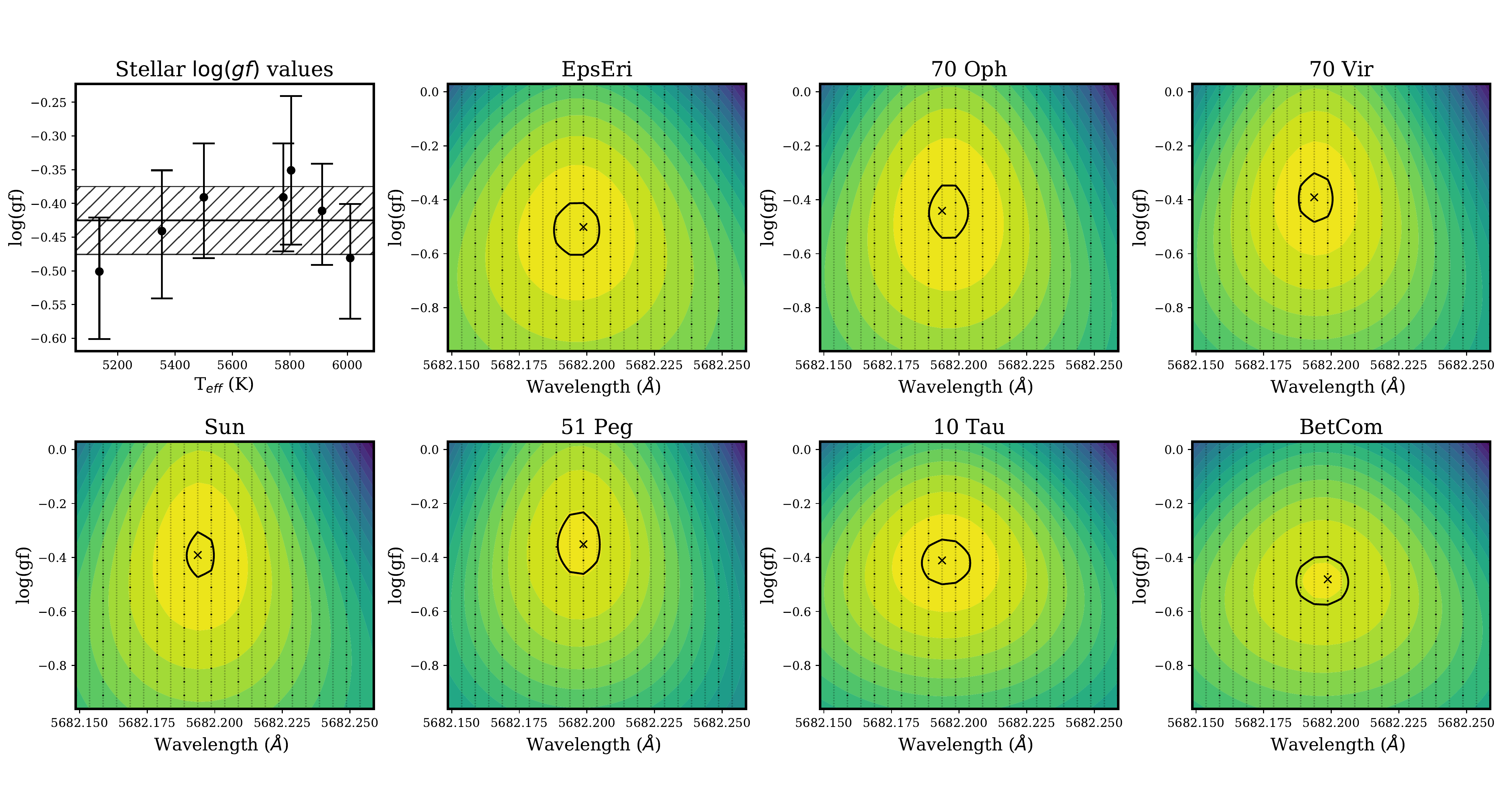}
\caption{ $\log(gf)_{grid}$ values and $\chi^2_{red}$ distributions for the $\lambda$5682.2 \ion{Ni}{I} line. Top left panel: Individual $\log(gf)_{grid}$ values per star, including 68.3\% confidence-limit error bars, plotted against benchmark $T_{\mathrm{eff}}$. The solid black line denotes the weighted $\overline{\log(gf)}_{grid}$ and the hatched region denotes the standard deviation estimation, discussed in Section~4.3.8. Remaining panels: $\chi^2_{red}$ distributions of the synthesised grid of $\lambda$ and $\log(gf)$ values for the seven benchmark stars. Calculated grid points are denoted by black dots, the individual $\log(gf)_{grid}$ values are denoted by black crosses, and the 68.3\% confidence limits are shown with black contours. Each plot has a normalised $\chi^2_{red}$ colour-scale, where yellow represents the $\chi^2_{red}$ minima, and dark blue represents the $\chi^2_{red}$ maximum.}
\label{gridmethod}
\end{figure*}

\subsection{Determination of wavelengths and oscillator strengths via iterative detailed modelling}

 Unlike the curve of growth line profiles, the grid spectra are computed using the line of interest and all other nearby lines present in the input atomic and molecular line lists.
A grid of synthetic spectra is calculated for each selected line per benchmark star, in a wavelength window of $\pm$1~\AA\ centred on the rest wavelength of the line. The grid method uses the rest-wavelength and the previously determined $\overline{\log(gf)}_{cog}$ values of the selected line as a central grid point, and covers the parameter ranges of -0.05~<~$\lambda_{input}$~<~+0.05~\AA\ (in steps of 0.01~\AA) and -0.5~<~$\log(gf)$~<~+0.5~dex (in steps of 0.05~dex). An example of the calculated grid line profiles is shown in Fig.~\ref{gridprofiles} for the central and extrema wavelength grid values.

The observed benchmark spectra and calculated line profiles are used to calculate a corresponding grid of $\chi^2_{red}$ values, which are interpolated in steps of $\Delta\lambda$~=~0.005~\AA\ and $\Delta\log(gf)$~=~0.01~dex using a bivariate cubic spline fit. The smallest $\chi^2_{red}$ value of the interpolated grid provides the final set of wavelength and $\log(gf)_{grid}$ values for the given line, for a given star, and confidence intervals at a 68.3\% confidence limit are calculated for the pair of parameters to produce upper and lower error estimates on the parameters. These individual wavelengths, $\log(gf)_{grid}$ values, and errors per star can be combined to produce a final mean wavelength, mean $\overline{\log(gf)}_{grid}$ value, and errors for the line (see Section~4.3.8). Figure~\ref{gridmethod} shows the $\chi^2_{red}$ grids calculated for the $\lambda$5682.2 \ion{Ni}{I} line in the several benchmark stars, in addition to the $\lambda_{grid}$ and $\log(gf)_{grid}$ values corresponding to the $\chi^2$ minimum per star, 68.3\% confidence limit error bars, and the $\overline{\log(gf)}_{grid}$ value of the line.

\subsection{Uncertainties in the astrophysical oscillator strengths }

\subsubsection{Uncertainties due to S/N}

\cite{cayrel1988} provides a formula for estimating the statistical uncertainty in measuring the $W_{\lambda}$ of an observed profile using a Gaussian fit. Using Eq. (7) of their work, the HERMES FWHM$\sim$0.065~\AA, a pixel size of $\delta x\sim$0.026~\AA, and a  S/N$\approx$800, a statistical uncertainty of $\sim$0.08~m\AA\ is determined for the equivalent width measurements. This statistical uncertainty only yields differences in $\log(gf)$ values on the order of the fourth decimal place, and so is negligible compared to the other sources of uncertainty.

\subsubsection{Uncertainties due to modelling assumptions}

We do not attempt an exhaustive investigation into all aspects of spectral modelling, such as adopting different model atmospheres and transfer codes, however we have adopted stellar objects for which the modelling uncertainties should be minimal. The plane-parallel atmospheric models of ATLAS9 were adopted, rather than the public MARCS models, as the MARCS model opacities do not contain enough ionised species to accurately model early-F stars and hotter \citep{plezmodels}, which is required for future plans to extend the analysis to hotter stars and ionised lines. The benchmark star selection mandated dwarf-like objects so that the plane-parallel assumption of ATLAS9 is valid. In the context of cool stars, \cite{gustafsson2008} show that the differences in temperature structures between ATLAS9 and MARCS are negligible for G-type dwarf stars of solar metallicity. Differences between ATLAS9, which uses ODFs, and ATLAS12, using opacity sampling, are shown to be extremely small when computed with the same chemical composition and line data \citep{plezmodels}.

As shown in \cite{lind2012}, the magnitude of non-LTE effects on unsaturated, high-excitation \ion{Fe}{I} lines for stars with stellar parameters of 5000~$\leq T_{\mathrm{eff}} \leq 6000$~K, $4.0 \leq \log{g} \leq 4.5$, and [M/H]~$\approx$~[M/H]$_{\odot}$ are between 0.00~and~0.02~dex at most,  and therefore the majority of investigated lines should be free of non-LTE effects. We do however expect the adopted stellar model atmospheres to begin to suffer from shortcomings for the hottest A-type benchmark stars, and the resultant non-LTE effects will lead to systematic errors in astrophysical $\log(gf)$ values. It is hard to predict the exact magnitudes of non-LTE effects for individual spectral lines, but lines of lower excitation energies belonging to easily ionised species are more likely to exhibit non-LTE effects. As discussed in Section 5.2, we do not find any obvious evidence of non-LTE effects for the majority of our astrophysical $\overline{\log(gf)}$ values, except for transitions belonging to the lowest energy levels of E$_{low} \leq 0.25$~eV.

\subsubsection{Uncertainties due to stellar parameters}

Differences between the physical stellar parameters and the adopted stellar parameters will impact the derived $\log(gf)$ values, as any difference in number densities will be attributed to the derived $\log(gf)$ values. We also assume that the benchmark stars have the same elemental abundance distribution as the Sun, but scaled with metallicity. This means that any changes to the relative abundance ratios of an individual element, for a given benchmark star, will manifest as $\Delta\log(gf)$ values for all lines belonging to that element. Due to the highly degenerate relationships between the stellar parameters, individual elemental abundances, and their dependence on the adopted atomic data, we do not propagate the stellar parameter errors into our $\log(gf)$ derivations or attempt to determine individual abundances. An accurate determination of these uncertainties warrants its own extensive investigation. Instead we adopt the position that the stellar parameters and metallicity scaled solar abundance ratios are correct, and that using multiple benchmark stars in the $\overline{\log(gf)}$ derivations makes the final $\overline{\log(gf)}$ values less susceptible to an erroneous set of stellar parameters or {changes in abundance ratios}. The standard deviation associated with the $\overline{\log(gf)}$ value of a line will reflect the uncertainties of any of the sets of stellar parameters or abundance ratios.

\subsubsection{Uncertainties due to normalisation}

The benchmark spectra presented in this work are normalised using an automatic template normalisation in order to prevent the introduction of `human factors' into the individual stellar spectra. The result should be a more consistent normalisation across the spectra, leading to a decrease in scatter, or random error, in the derived $\log(gf)$ values of a given line. To account for any deviations from unity in local continuum placement, the Gaussian $W_{\lambda}$ measurements are performed with a flat continuum offset term. In practice, this offset term allows the single Gaussian fit to mitigate the influence of nearby broad-winged lines. The iterative fitting does not attempt to measure any deviations in local continuum placement.

\begin{figure}
\centering
\includegraphics[width=9cm]{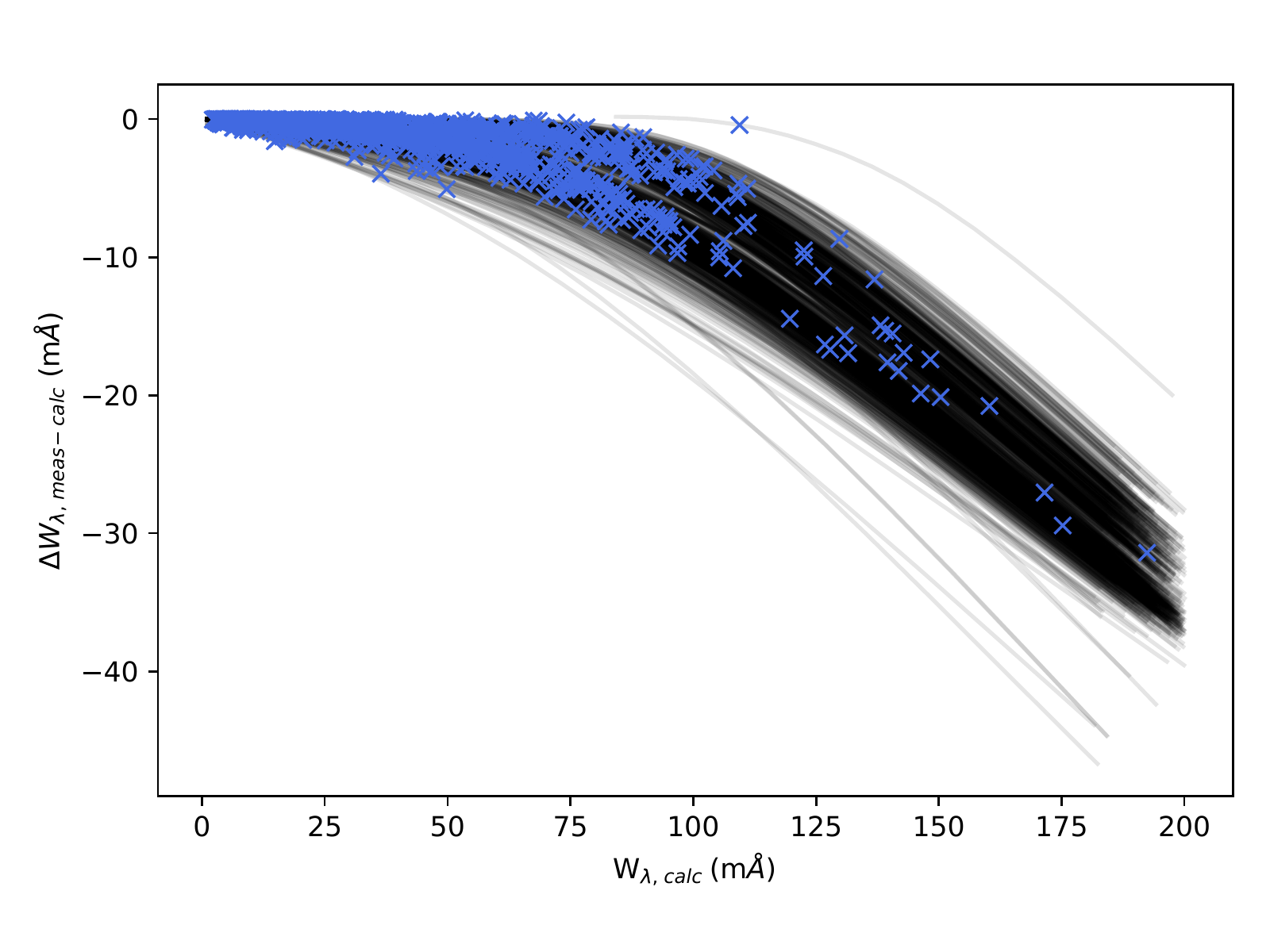}
\caption{Predicted $\Delta W_{\lambda}$ between Gaussian fits and transfer-calculation Voigt profiles plotted as a function of calculated $W_{\lambda}$ shown for the 1091 investigated lines in the Sun (black lines). Blue crosses denote the proposed $\Delta W_{\lambda}$ corrections for the 1091 spectral lines. }
\label{gaussian_departure}
\end{figure}

\begin{figure}
\centering
\includegraphics[width=9cm]{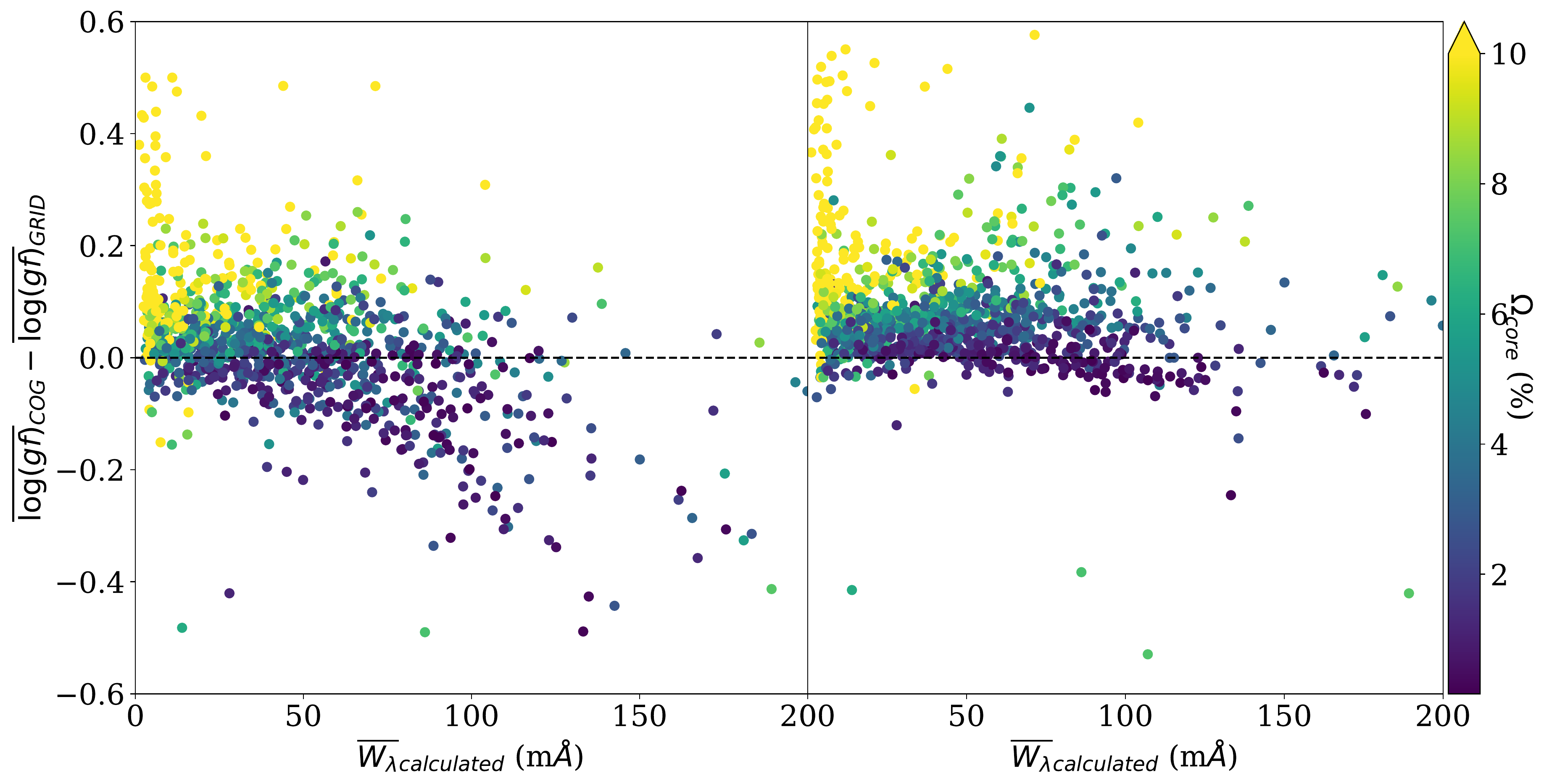}
\caption{Left: $\Delta\overline{\log(gf)}_{cog-grid}$ plotted against the calculated $\overline{W_{\lambda}}$ prior to the corrections to the Gaussian profiles. Right: As in \textit{Left} but after corrections are applied on a line-per-line, star-by-star basis. The colour scale corresponds to the mean $\Omega_{core}$ (in percentage) of the line in question, recalculated using the new $\overline{\log(gf)}_{cog}$ values. The stratification of the $\Omega_{core}$ is explained in Section~4.3.7.
 }
\label{gaussian_correction}
\end{figure}

\begin{figure*}
\centering
\includegraphics[width=18.5cm]{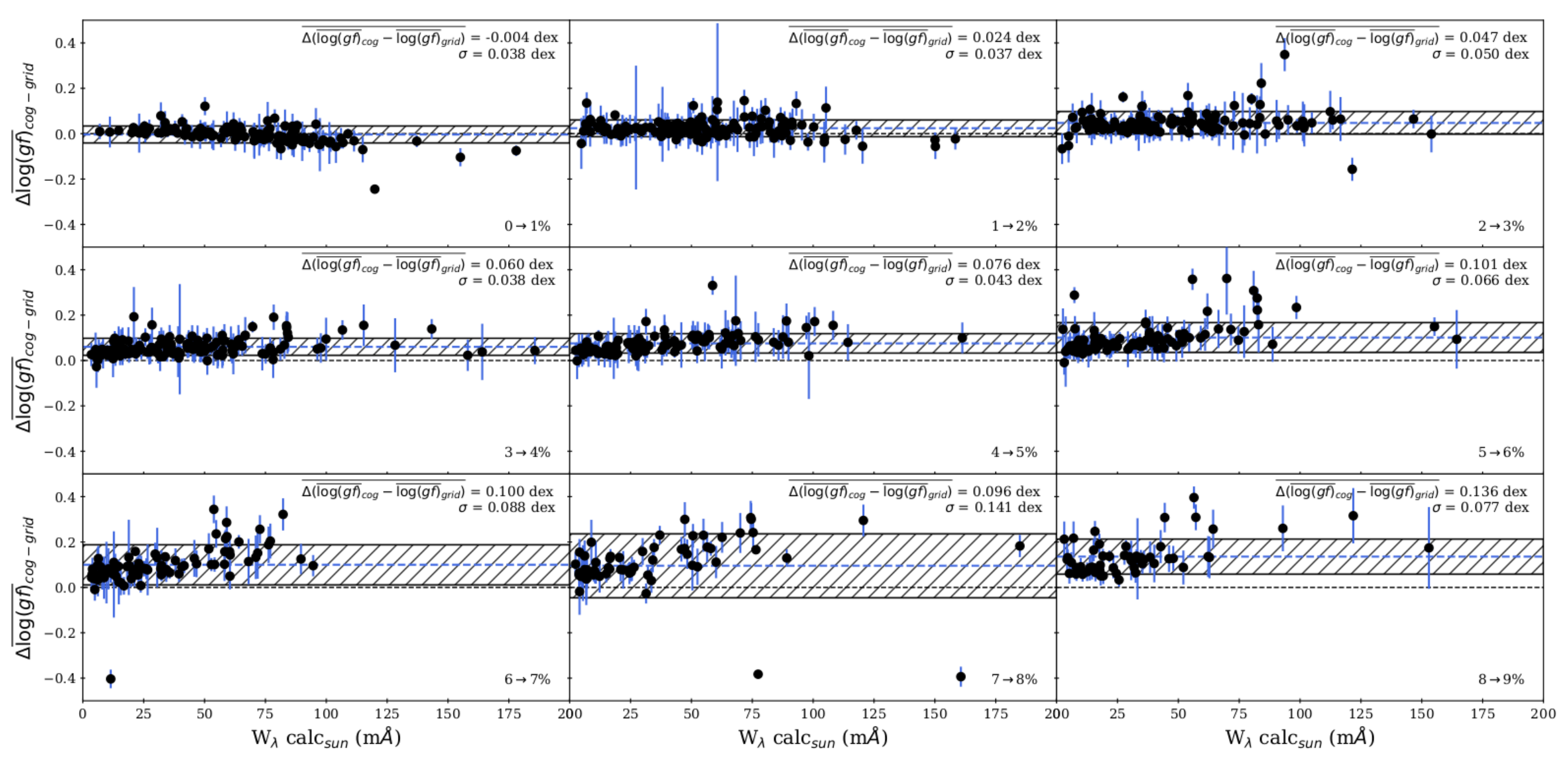}
\caption{$\Delta\overline{\log(gf)}_{cog-grid}$ plotted against theoretical $W_{\lambda}$ values for the Sun, shown for different steps of $\Omega_{core}$ (left to right, top to bottom). The mean difference of the $\Delta\overline{\log(gf)}_{cog-grid}$ values is shown by the dashed blue lines, and the standard deviation is denoted by the hatched area.    }
\label{blending_trend}
\end{figure*}

\begin{figure}
\centering
\includegraphics[width=9cm]{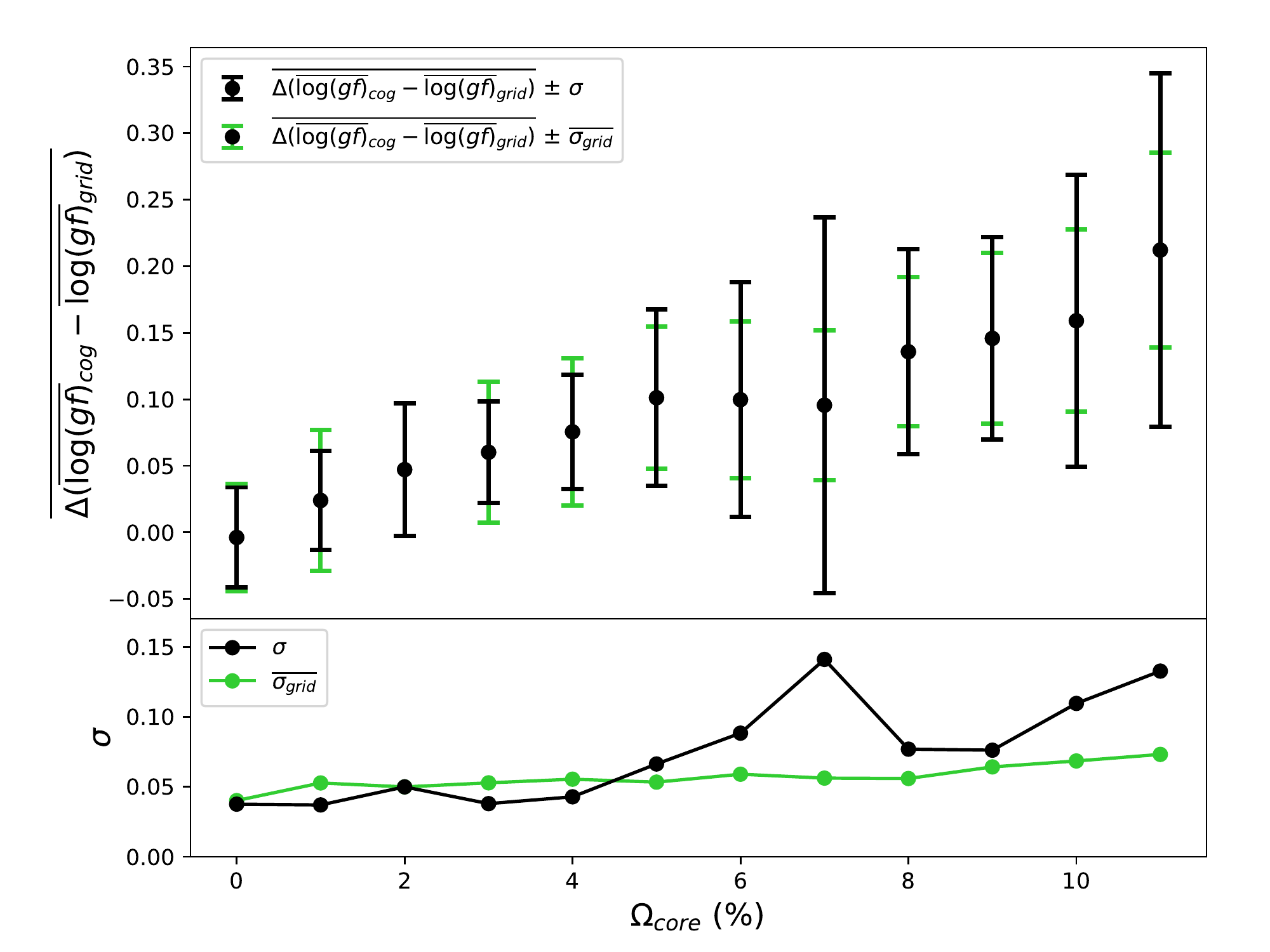}
\caption{Top panel: The mean $\Delta\overline{\log(gf)}_{cog-grid}$ values plotted against $\Omega_{core}$. Black error bars show the standard deviation $\sigma$ of the $\Delta\overline{\log(gf)}_{cog-grid}$ values, and green error bars show mean of the $\sigma_{grid}$ values. Lower panel:  $\sigma$ and $\overline{\sigma}_{grid}$ values plotted as a function of $\Omega_{core}$. The mean of the $\Delta\overline{\log(gf)}_{cog-grid}$ values is shown to increase as $\Omega_{core}$ increases. The $\sigma$ also increases with $\Omega_{core}$, however the $\overline{\sigma}_{grid}$ exhibits a flatter trend. At $\Omega_{core} \leq$~5\%, $\sigma$ is almost constant suggesting that the intrinsic scatter between the two methods is $\sigma = 0.04$~dex. }
\label{blending_trend_sigma}
\end{figure}

\subsubsection{Impact of Gaussian profile for $W_{\lambda}$ measurements}

Spectral lines are often approximated as Gaussian profiles rather than Voigt profiles, however this assumption breaks down as lines become saturated and the profile wings begin to grow. Fitting a line with a Gaussian profile will lead to an underestimation of the line $W_{\lambda}$  as a function of increasing $W_{\lambda}$. This issue is discussed by \cite{luck2017}, in addition to the difficulties of performing physically accurate Voigt profile fits automatically for many lines.

In order to theoretically quantify this underestimation, Gaussian fits are performed on all synthetic line profiles of the curves of growth calculated in Section~4.1. Figure~\ref{gaussian_departure} shows the difference in $W_{\lambda}$ between the two profile shapes as a function of $W_{\lambda}$, computed for the 1091 lines in the Solar spectrum. The $W_{\lambda}$ difference was calculated on a line-by-line, star-by-star basis and applied to each of the corresponding $W_{\lambda}$ measurements discussed in Section~3.2. Figure~\ref{gaussian_correction} shows the difference between the $\overline{\log(gf)}_{cog}$ values, measured using a Gaussian profile fit, and  $\overline{\log(gf)}_{grid}$ values, calculated using the full radiative-transfer broadening, before and after the correction to the $W_{\lambda}$ values. After correction there is a distinct improvement in the observed negative trend between $\Delta\overline{\log(gf)}_{cog-grid}$ and $W_{\lambda}$, with the remaining positive scatter caused by differences in the treatment of theoretical blending between the two methods, as discussed in Section~4.3.7.

\subsubsection{Uncertainties due to line blending in observed spectra}

Observable blends,  that is, those that exhibit clear asymmetries or multiple peaks, are manually flagged and excluded on a line-by-line, star-by-star basis, to mitigate their impact on the $\log(gf)$ derivations. Hidden blends, that is, those that do not cause measurable asymmetries or additional peaks in the line profile of interest, will cause an overestimation of the derived $\log(gf)$ value unless there is prior theoretical knowledge of this blend (see Section~4.3.7). If the two blending lines behave differently with temperature then it is expected that the benchmark stars will produce systematic differences in the derived $\log(gf)$ value. As such, the standard deviation of the $\log(gf)$ values for a given line can be used as an indicator as to whether or not a line may be affected by hidden blends, and thus whether or not it is reliable to use as a single unblended line in our analysis and quality assessment for this paper.

\subsubsection{Uncertainties due to theoretical line blending}

The curve of growth approach assumes that a measured spectral feature is completely unblended, attributing the measured $W_{\lambda}$ to a single spectral line, whereas the iterative modelling approach may contain some knowledge of potential blending components in the input line list. This difference will lead the curve of growth approach to overestimate the $\log(gf)_{cog}$ of the line of interest relative to the iterative modelling approach, assuming that the blending component is actually present in the observed spectrum.

Figure~\ref{blending_trend} shows $\overline{\log(gf)}_{cog} - \overline{\log(gf)}_{grid}$ plotted against the theoretical $W_{\lambda}$ of the Sun, for intervals of $\Omega_{core}$ that have been recalculated using the new $\overline{\log(gf)}$ values. Figure~\ref{blending_trend_sigma} shows the mean $\Delta\overline{\log(gf)}_{cog-grid}$ plotted as a function of $\Omega_{core}$, as well as the standard deviation of the $\overline{\log(gf)}_{cog-grid}$ values and $\overline{\sigma}_{grid}$ values plotted as functions of $\Omega_{core}$. The mean $\Delta\overline{\log(gf)}$ is 0.00~dex when $\Omega_{core} \ll$~1\%, and increases steadily with $\Omega_{core}$, confirming that the systematic difference between the methods is due to theoretical blending. The standard deviation similarly increases with $\Omega_{core}$, but levels out at roughly $\Omega_{core} \leq$~5\% at a value of 1$\sigma=\pm$0.04~dex, which therefore seems to represent the intrinsic scatter between the two methods. Properly correcting for this difference in methods is difficult as one must know how the background line presence would affect the single Gaussian fit, and rely on the assumption that the background line atomic data are accurate. Given that properly rectifying this difference would require one to simultaneously quality assess two or more blended lines, we instead recommend using the intrinsic scatter value of 0.04~dex as a constraint on line selection, in order to reduce the impact of this systematic difference on the quality assessment results (see Section 5.1).

\subsubsection{Final $\overline{\log(gf)}$ values and uncertainties}

The $\overline{\log(gf)}_{cog}$ values are calculated by taking the mean of the \textit{n} individually derived stellar $\log(gf)_{cog}$ values, after individual corrections for the Gaussian approximation discussed in Section~4.3.5. The standard deviation of the ${\log(gf)}_{cog}$ values, $\sigma_{cog}$, are calculated as shown in Eq.~4:

\begin{equation}
\begin{gathered}
{\sigma}_{cog} = \sqrt{ \dfrac{1}{n - 1} \sum_{i=1}^{n}  \left ( \log(gf)_{i} - \overline{\log(gf)}_{cog} \right ) ^2 }
\end{gathered}
.\end{equation}

The $\overline{\log(gf)}_{grid}$ values are calculated by using a weighted mean of the \textit{n} individually derived stellar $\log(gf)_{grid}$ values, shown by Eq.~5, and the corresponding standard deviation estimations are shown by Eqs.~6~and 7, where the weights per star are equal to the inverse square of the $\log(gf)_{grid}$ uncertainty estimates:

\begin{equation}
\begin{gathered}
\overline{\log(gf)}_{grid} = \dfrac{\sum_{i=1}^{n} \left ( \log(gf)_{i} / \sigma_i^2 \right ) }{\sum_{i=1}^{n} 1/\sigma_i^2 }
\end{gathered}
,\end{equation}
and 
\begin{equation}
\begin{gathered}
{\sigma}_{grid}  = \sqrt{    \dfrac{\sum_{i=1}^{n} \left (  \log(gf)_i - \overline{\log(gf)}_{grid}     \right )^2 / \sigma_i^2   }{
               V_1 - ( V_2/ V_1 ) }  }
\end{gathered}
,\end{equation}
where V$_1$ and V$_2$ are defined as follows:
\begin{equation}
\begin{gathered}
V_1  = \sum_{i=1}^{n} 1/\sigma_i^2  \qquad  V_2  = \sum_{i=1}^{n}  \left( 1/\sigma_i^2 \right)^2
\end{gathered}
.\end{equation}

In the case that all the weights are equal, Eqs.~6~and~7 simplify to take the same form as the standard deviation shown in Eq.~4.

\section{Quality assessment and discussion}

\begin{figure}
\centering
\includegraphics[width=9cm]{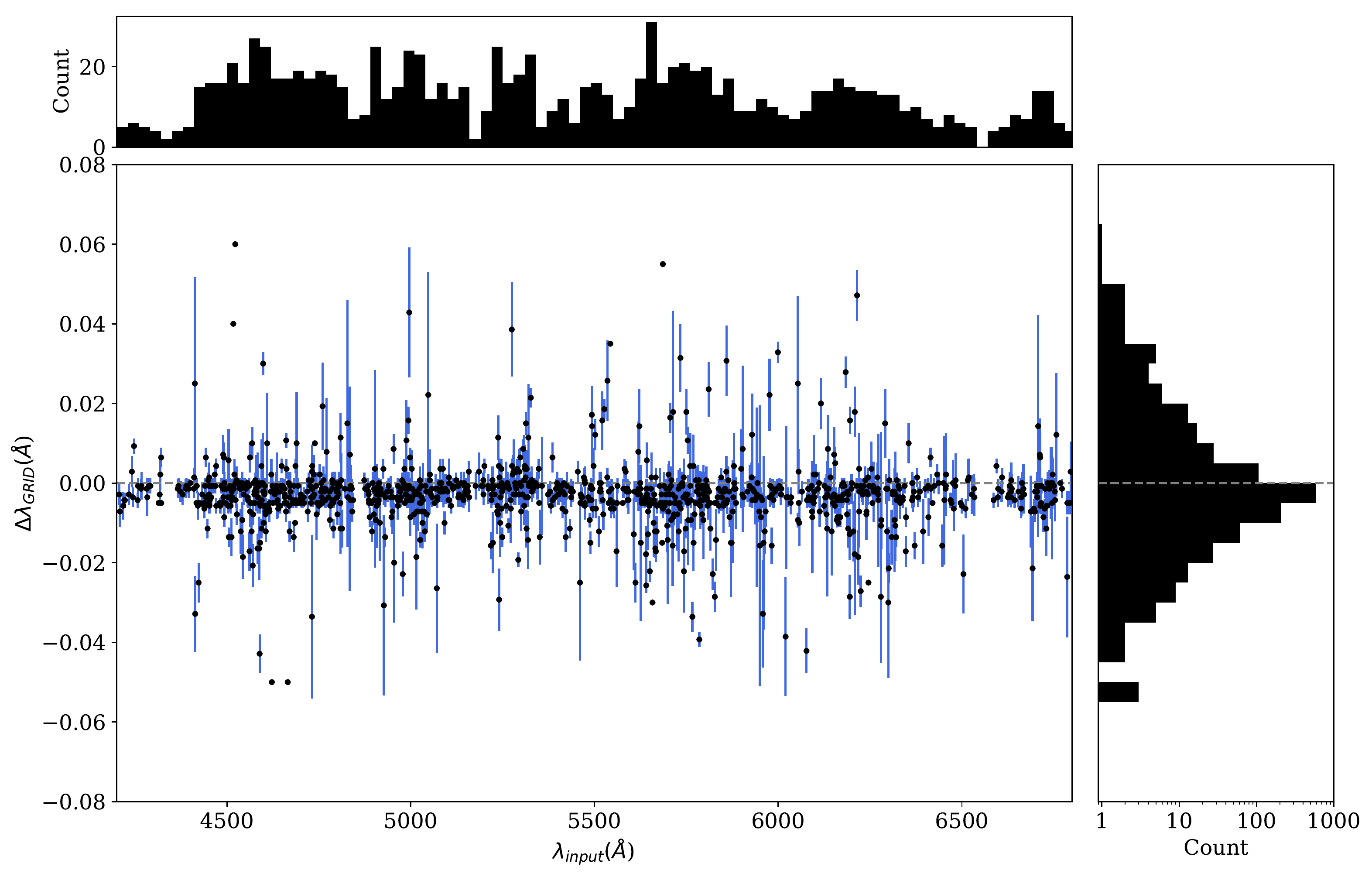}
\caption{$\Delta\overline{\lambda}_{grid-input}$ vs. $\lambda_{input}$ for the 1091 investigated lines. 1$\sigma$ error bars are shown in blue. A slight negative offset is present, however this is within the uncertainty in the HERMES spectrograph wavelength calibration. A small fraction of transitions have wavelength offsets that are attributed to inaccuracies in the atomic data. }
\label{wav_scatter}
\end{figure}

\begin{figure*}
\centering
\includegraphics[width=16cm]{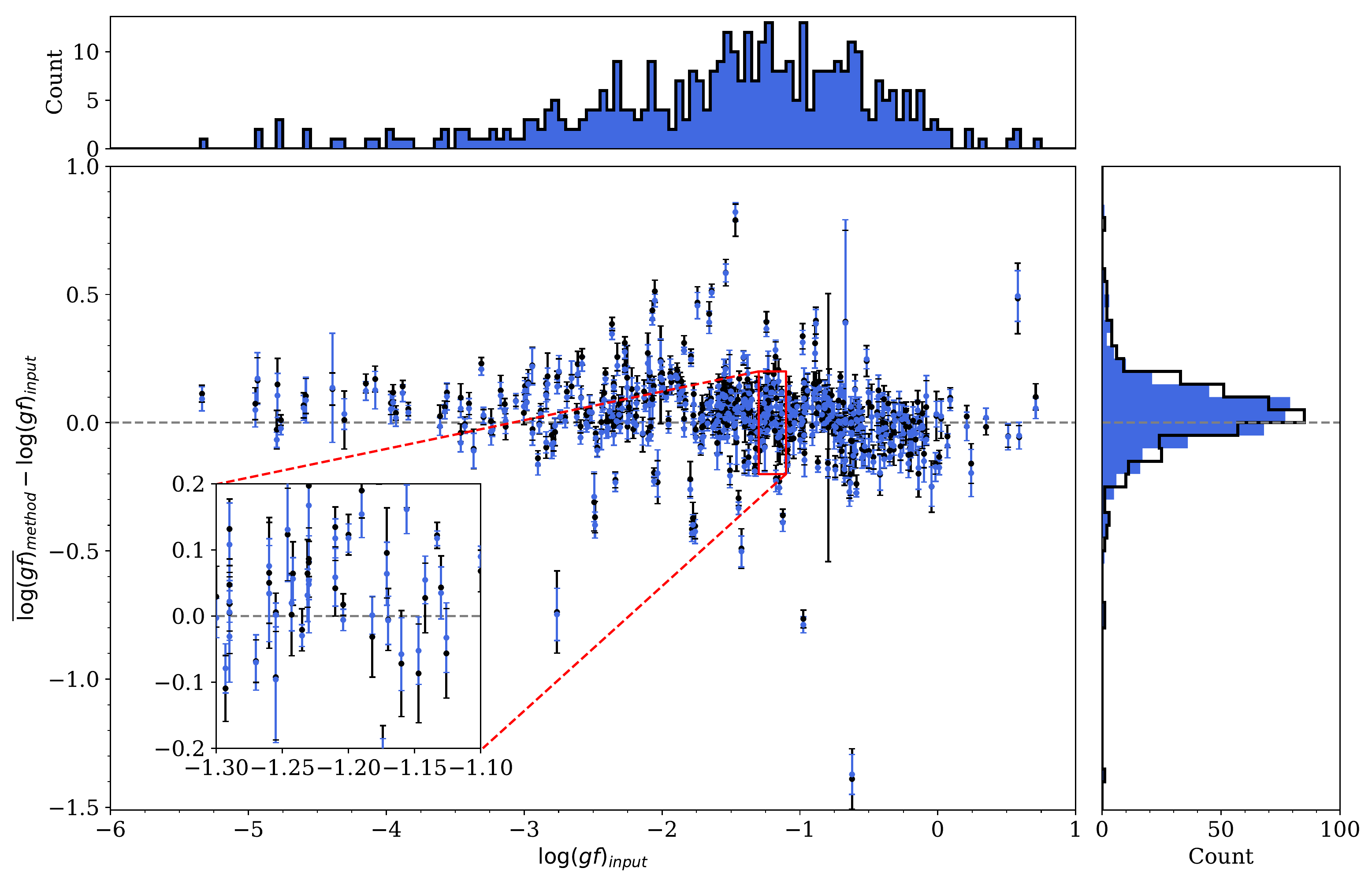}
\caption{$\Delta\overline{\log(gf)}_{grid-input}$ (in blue) and $\Delta\overline{\log(gf)}_{cog-input}$ (in black) plotted against $\log(gf)_{input}$ for the 408 analysis-independent lines discussed in Section~5.1. There are no clear trends in either the $\Delta\overline{\log(gf)}_{grid-input}$ values or $\Delta\overline{\log(gf)}_{cog-input}$ values with respect to $\log(gf)_{input}$, however there are a significant number of transitions for which both methods, which agree within 0.04~dex of each other, find sizeable offsets compared to the $\log(gf)_{input}$. Some transitions even show differences between $\log(gf)_{input}$ and our $\overline{\log(gf)}$ values of $\pm 0.50$~dex or more. The zoomed subplot shows transitions that have significantly smaller differences between the derived and input $\log(gf)$ values, however a number of them still do not agree within the derived uncertainties.
}
\label{loggf_scatter_robust}
\end{figure*}

\begin{figure}
\centering
\includegraphics[width=9cm]{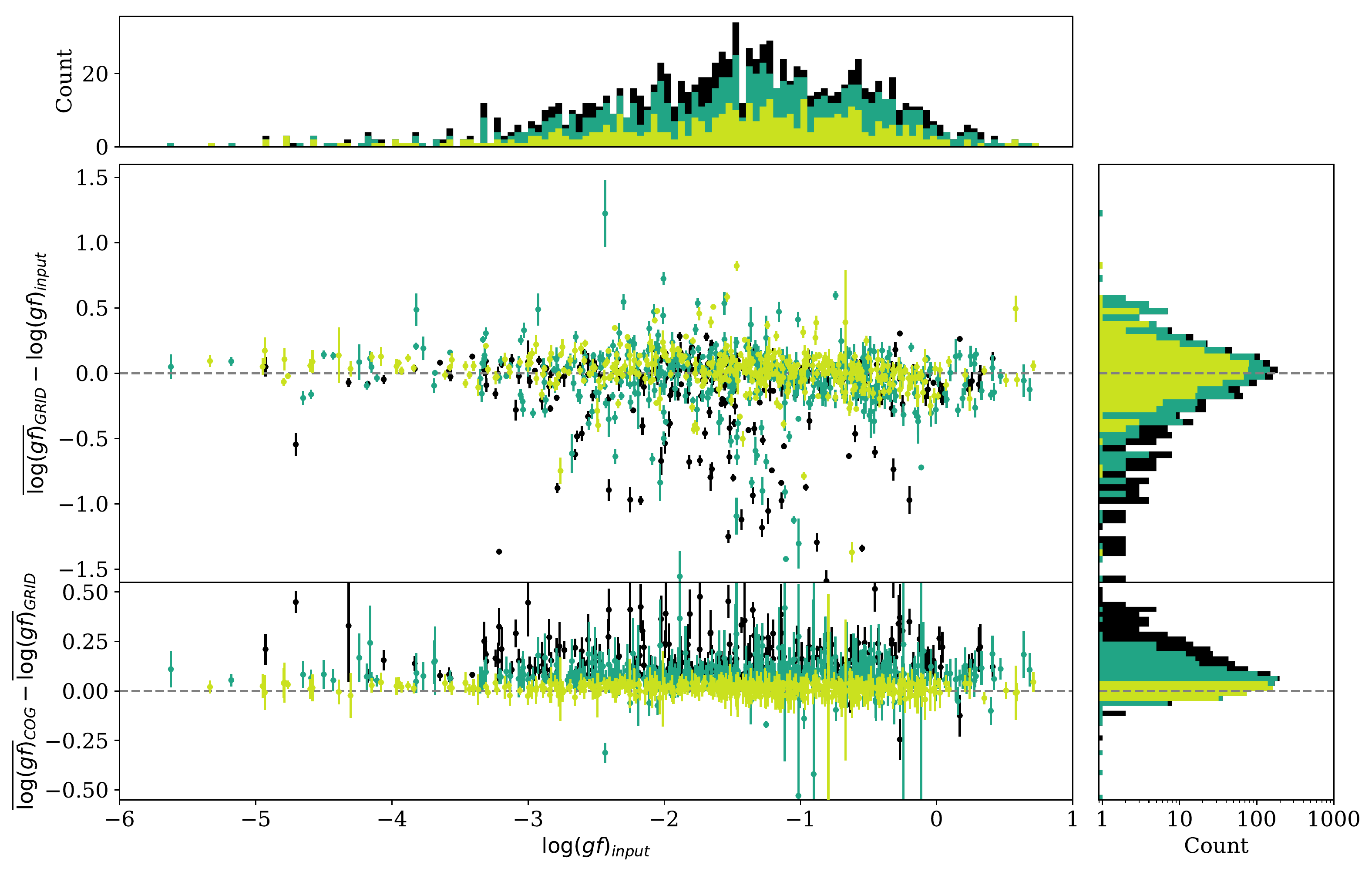}
\caption{$\Delta\overline{\log(gf)}_{grid-input}$ vs. $\log(gf)_{input}$ for the 1091 investigated lines.  Bottom panel: As above but for $\Delta\overline{\log(gf)}_{cog - grid}$. The 408 analysis-independent lines are shown in light green, the remainder of the 845 quality-assessable lines are shown in dark green, and the remainder of the 1091 investigated lines are shown in black. There are no obvious trends in either $\Delta\overline{\log(gf)}_{grid-input}$ or $\Delta\overline{\log(gf)}_{cog-grid}$ with respect to $\log(gf)_{input}$, except for a number of transitions where both methods, agreeing with each other   within errors, find significant upward revisions of the $\log(gf)_{input}$ values as large as $\Delta\log(gf) = 1.5$~dex. 
}
\label{loggf_scatter}
\end{figure}

\begin{figure*}
\centering
\includegraphics[width=16cm]{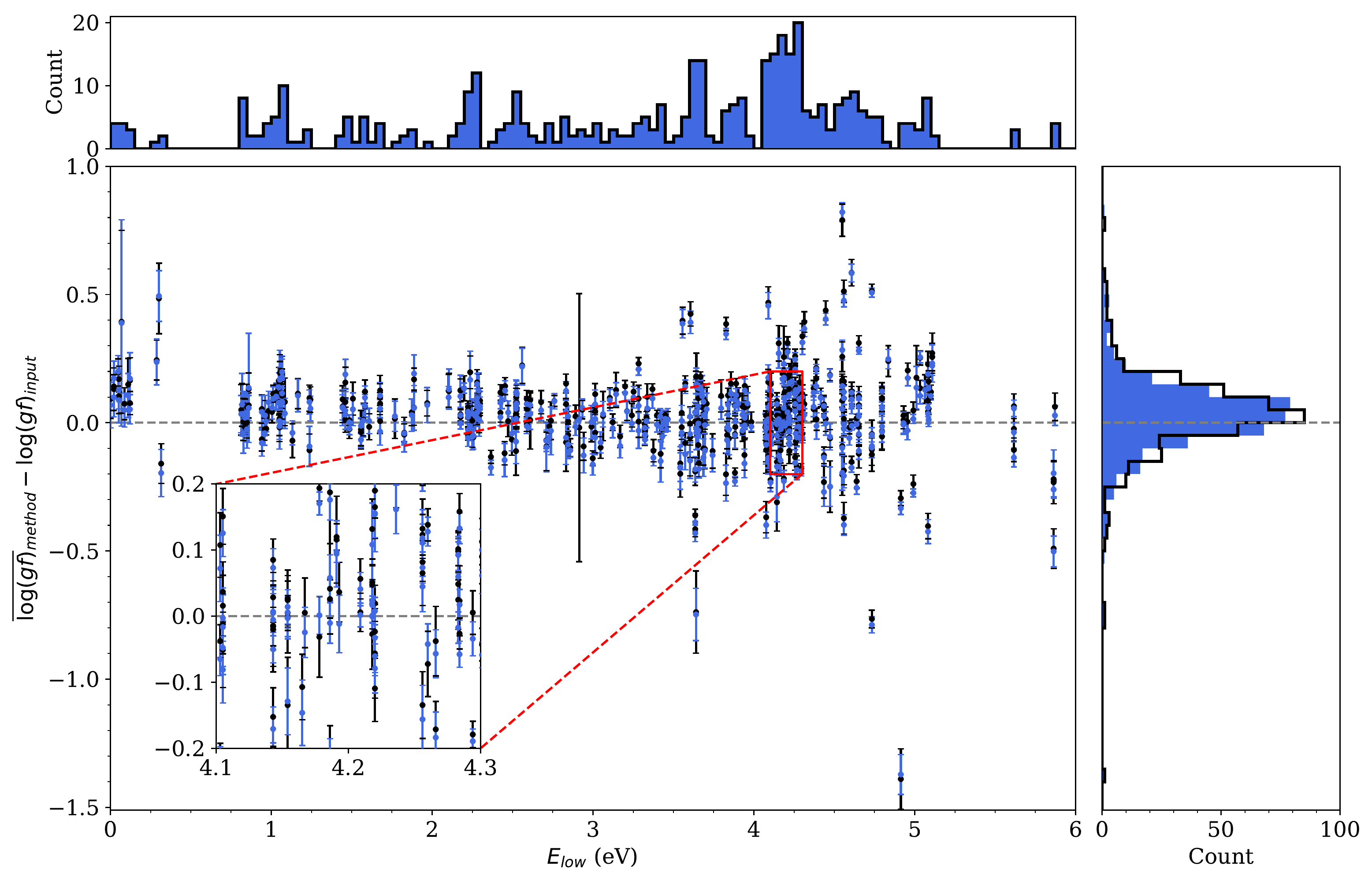}
\caption{As in Fig.~\ref{loggf_scatter_robust}, but plotted as a function of E$_{low}$. There is a slight upward inflection of $\overline{\log(gf)}$ values at excitation energies of E$_{low} \leq 0.25$~eV which is possibly a symptom of non-LTE effects in these low-level transitions. Between 3.5~$\leq$~E$_{low}$~$\leq$~5.0~eV we find an increase in scatter of $\overline{\log(gf)}$ values, often belonging to higher excitation \ion{Fe}{I} transitions. The zoomed subplot shows that a significant number of analysis-independent transitions have multiplets in common, and that the $\log(gf)$ values of these multiplets can show significant scatter. 
}
\label{loggf_elow_robust}
\end{figure*}

\begin{figure}
\centering
\includegraphics[width=9cm]{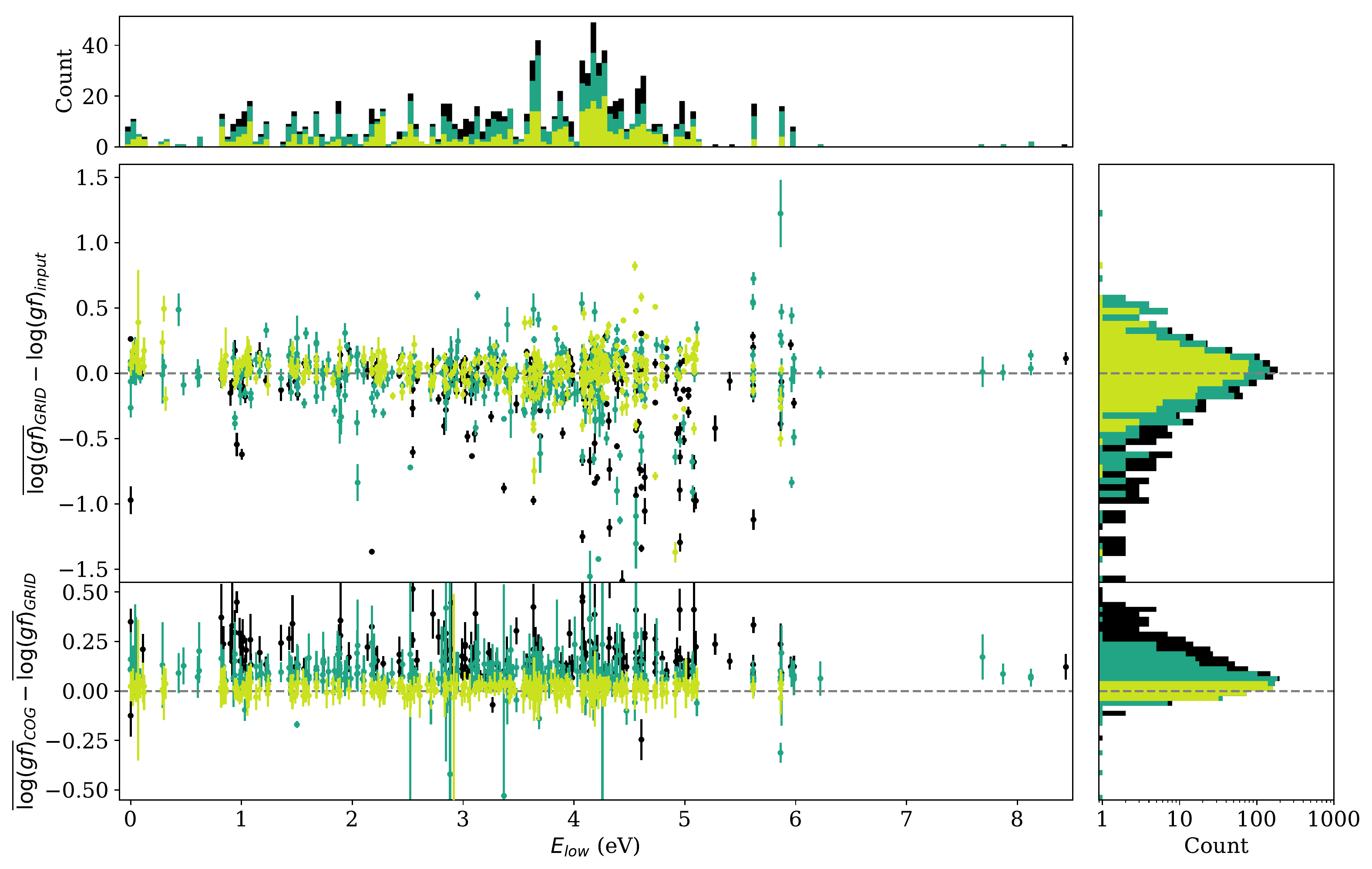}
\caption{As in Fig.~\ref{loggf_scatter}, but plotted as a function of E$_{low}$. The large downward revisions pertain to \ion{Fe}{I} transitions originating from higher excitation energies. The upward trend towards E$_{low}$~=~0~eV is still visible, however to a lesser degree than the distribution of solely the analysis-independent lines.
}

\label{loggf_elow}
\end{figure}

\begin{figure}
\centering
\includegraphics[width=9cm]{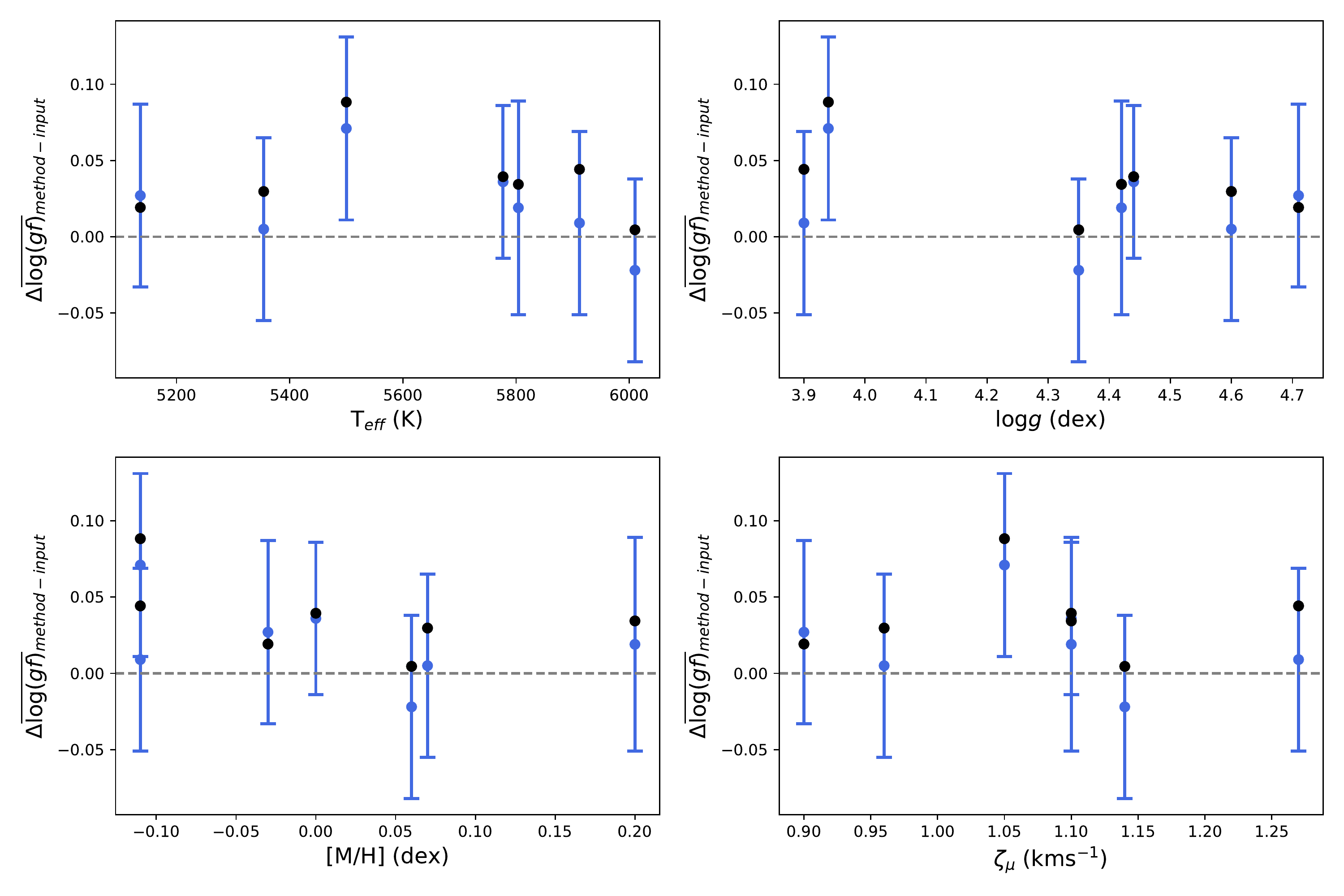}
\caption{ Median of the 408 analysis-independent $\Delta\overline{\log(gf)}_{method-input}$ values, for both the curve of growth (black circles) and iterative modelling (blue circles) methods, are plotted against the stellar parameters of the seven benchmark stars. The median of the $\sigma_{grid}$ values are denoted by the blue error bars. The overestimation of the $\overline{\log(gf)_{cog}}$ relative to the $\overline{\log(gf)_{grid}}$ values can be seen. Apart from \ion{V}{I}, we do not find any clear trends with respect to the stellar parameters for any of the 14 remaining species in the analysis-independent subset of lines. 
}
\label{loggf_AP}
\end{figure}

\subsection{Selection of quality-assessable lines}
A line is considered quality-assessable if both the curve of growth and iterative methods derive $\overline{\log(gf)}$ values with overlapping errors, otherwise the line is deemed too difficult to accurately work with, be it due to measuring, modelling, or blending issues. An additional constraint, requiring the $\Delta\overline{\log(gf)}$ of the two methods to be $\leq$~0.04~dex, can be imposed to increase the likelihood that the selected lines are free of systematic differences in analysis methods due to theoretical blending. Lines that fulfil this criterion are referred to as {analysis-independent} lines. As an example of the quality-assessable line selection, Figs.~\ref{invest_example},~\ref{qa_example}, and~\ref{robust_example} of the Appendix show the synthesised line profiles belonging to the $\overline{\log(gf)}_{cog}$ and $\overline{\log(gf)}_{grid}$ values, plotted against the observed line profile, for a non-quality-assessable line (Fig.~\ref{invest_example}), a quality-assessable line (Fig.~\ref{qa_example}), and an analysis-independent spectral line (Fig.~\ref{robust_example}), for the several benchmark stars. The quality of the observed data is sufficiently high that even lines as weak as the central~line~depth $d \approx 0.02$ can be reliably quality-assessed.

\subsection{Distributions of derived $\lambda$ and $\log(gf)$ values}
Figure~\ref{wav_scatter} shows the $\Delta\overline{\lambda}_{grid-input}$ plotted against $\lambda_{input}$ for the 1091 investigated lines. For most lines there is excellent agreement between the literature wavelengths and those derived via the iterative modelling. A small fraction of transitions have significant offsets of up to $|\Delta\lambda| \leq$~0.06~\AA, which are caused by inaccuracies in the atomic data. There is a slight negative offset in $\Delta\overline{\lambda}_{grid-input}$, however this is well within the uncertainties of the HERMES spectrograph wavelength calibration.

Figure~\ref{loggf_scatter_robust} shows the $\Delta\overline{\log(gf)}_{grid-input}$ and $\Delta\overline{\log(gf)}_{cog-input}$ plotted against $\log(gf)_{input}$ for the 408 analysis-independent lines. We find that for a number of transitions both methods show significant revisions of the $\log(gf)_{input}$ values as large as $\pm 0.50$~dex or more. We do not find any obvious trends between the $\log(gf)$  and $\log(gf)_{input}$ values resulting from the two methods, nor is there a correlation between the uncertainties and either (a) the $\Delta\log(gf)_{method-input}$ values or (b) the $\log(gf)_{input}$ values. Figure~\ref{loggf_scatter} shows the $\Delta\overline{\log(gf)}_{grid-input}$ plotted against $\log(gf)_{input}$ for the 1091 investigated lines, in addition to the $\overline{\log(gf)}_{cog-grid}$ plotted against $\log(gf)_{input}$. Both the 845 quality-assessable lines and the 1091 investigated lines contain a number of transitions for which we find a significant downward $\log(gf)$ revision of up to 1.50~dex.  

Figure~\ref{loggf_elow_robust} shows the $\Delta\overline{\log(gf)}_{grid-input}$ and $\Delta\overline{\log(gf)}_{cog-input}$ plotted against $\log(gf)_{input}$ for the 408 analysis-independent lines. We find an increase in scatter of the derived $\log(gf)$ values for higher excitation transitions. This increase in scatter is uncorrelated with the uncertainties of the derived $\log(gf)$ values, and is emphasised even more by the quality-assessable lines shown in Fig.~\ref{loggf_elow}. We also find a positive offset for transitions belonging to the lowest excitation energies. This is possibly due to non-LTE effects in a small number of lines belonging to either the lowest terms of easily ionised species, or resonant lines of other species. Figure~\ref{loggf_elow} shows the $\Delta\overline{\log(gf)}_{grid-input}$ plotted against $E_{low}$ for the 1091 investigated lines. The increase in scatter of  $\Delta\overline{\log(gf)}_{grid-input}$ with increasing $E_{low}$ is present for the species \ion{Fe}{I} and \ion{Si}{I}, with the vast majority of the $\log(gf)_{input}$ values originating from the theoretical Kurucz 2007 line lists \citep{tableref15}. The likely cause for this apparent relationship is transitions with a high degree of atomic configuration mixing, for which $\log(gf)$ values are difficult to calculate.

Figure~\ref{loggf_AP} shows the median of the 408 analysis-independent $\Delta\overline{\log(gf)}_{method-input}$ values, for both the curve of growth and iterative modelling methods, plotted against the benchmark stellar parameters.  There is a small offset between the curve of growth and iterative modelling methods which is caused by the different treatments of line blending. There is no clear trend between the median $\Delta\overline{\log(gf)}_{method-input}$ values and any of the stellar parameters, with the median $\Delta\overline{\log(gf)}_{grid-input}$ values centred around $\sim$0.00~dex, suggesting that the analysis is not subject to significant systematic modelling issues with respect to the stellar parameters. The exception to this is \ion{V}{I}, which shows a negative trend with respect to increasing effective temperature. The cause of this is not clear, and could be linked to a number of possibilities: hyperfine splitting,  which can easily be seen in the profile shapes of the $\lambda$4379.2 and $\lambda$4594.1 \ion{V}{I} lines (see Appendix Table~\ref{appendixtable} hyperlinks for interactive line plots); non-LTE effects, as \ion{V}{I} is an easily ionised species and several of the \ion{V}{I} lines originate from low-lying energy levels; or even differences in the [V/Fe] ratio between the different benchmark stars. Given that the same trend appears in multiple \ion{V}{I} lines we believe the trend is less likely to be caused by line blending issues.

\begin{table}
\centering
\caption{ Elemental and ionic distribution of the investigated, quality-assessable, and analysis-independent lines. In brackets are the number of lines with atomic data within $\sigma_{grid}$ of the $\overline{\log(gf)}_{grid}$ values. It is worth noting that isotopic and hyperfine lines did not receive special treatment throughout the analysis, which may have lead to their under-representation in the line sets. }
\begin{tabular}{lccc}
            \hline
            \hline
            \noalign{\smallskip}
            Species      & number of  & quality & analysis  \\
            & investigated lines & assessable & independent \\
            \noalign{\smallskip}
            \hline
            \noalign{\smallskip}

                        \ion{C}{I}   & 1   & 1 \tiny{(1)}    & -\\
                        \ion{Na}{I}  & 3   & 3 \tiny{(1)}    & 2 \tiny{(0)}\\
                        \ion{Mg}{I}  & 5   & 5 \tiny{(2)}    & 2 \tiny{(0)}\\
                        \ion{Al}{I}  & 2   & 2 \tiny{(0)}    & 1 \tiny{(0)}\\
                        \ion{Si}{I}  & 56  & 42 \tiny{(19)}  & 14 \tiny{(7)}~~~\\
                        \ion{Si}{II} & 2   & 2 \tiny{(2)}   & -\\
                        \ion{S}{I}   & 1   & 1 \tiny{(1)}   & -\\
                        \ion{Ca}{I}  & 21  & 21 \tiny{(9)}~~~  & 10 \tiny{(4)}~~~\\
                        \ion{Ca}{II} & 1   & -   & -\\
                        \ion{Sc}{I}  & 1   & 1 \tiny{(1)}   & -\\
                        \ion{Sc}{II} & 19  & 17 \tiny{(12)}  & 3 \tiny{(2)}\\
                        \ion{Ti}{I}  & 90  & 80 \tiny{(60)}  & 47 \tiny{(32)}\\
                        \ion{Ti}{II} & 36  & 28 \tiny{(21)}  & 10 \tiny{(9)}~~~\\
                        \ion{V}{I}   & 15  & 14 \tiny{(12)}  & 6 \tiny{(4)}\\
                        \ion{V}{II}  & 1   & 1 \tiny{(1)}   & -\\
                        \ion{Cr}{I}  & 82  & 61 \tiny{(44)}  & 36 \tiny{(24)}\\
                        \ion{Cr}{II} & 12  & 7 \tiny{(3)}   & 3 \tiny{(1)}\\
                        \ion{Mn}{I}  & 13  & 9 \tiny{(7)}   & 1 \tiny{(1)}\\
                        \ion{Fe}{I}  & 543 & 401 \tiny{(146)} & 231 \tiny{(92)}~~~\\
                        \ion{Fe}{II} & 30  & 15 \tiny{(11)}~~~  & 4 \tiny{(4)}\\        
                        \ion{Co}{I}  & 15  & 9 \tiny{(6)}   & -\\
                        \ion{Ni}{I}  & 126 & 112 \tiny{(86)}~~~ & 38 \tiny{(30)}\\
                        \ion{Zn}{I}  & 2   & 2 \tiny{(0)}   & -\\
                        \ion{Sr}{I}  & 1   & 1 \tiny{(0)}   & -\\
                        \ion{Y}{II}  & 9   & 8 \tiny{(5)}   & -\\                       
                        \ion{Ba}{II} & 1   & -   & -\\
                        \ion{La}{II} & 2   & 1 \tiny{(1)}   & -\\
                        \ion{Ce}{II} & 1   & 1 \tiny{(0)}   & -\\

            \hline
            \noalign{\smallskip}
                        Total: & 1091 & 845 \tiny{(451)} & 408 \tiny{(210)}\\
            \hline            
\end{tabular}
\label{linedistribution}
\end{table}

\subsection{Discussion of quality assessment results}

 Of the 1091 atomic lines investigated, 845 are found to be quality-assessable and 408 of them are found to be analysis-independent lines. The ionic distribution of these lines is listed in Table~\ref{linedistribution}. Approximately half of the investigated and quality-assessable lines belong to \ion{Fe}{I}, and another  $\sim$10\% of the lines belong to singly ionised species. The literature atomic data belonging to a quality-assessable line are considered in agreement with our work, and therefore we can be recommend the value for use, if it agrees within the errors of the derived $\overline{\log(gf)}_{grid}$ and $\sigma_{grid}$ values. We do not make use of any literature errors as these are only available in a handful of the retrieved databases. The $\overline{\log(gf)}_{grid}$ value is adopted as the benchmark value as in almost all cases it produces better $\chi^2_{red}$ values than the $\overline{\log(gf)}_{cog}$ value. In addition, the $\overline{\log(gf)}_{cog}$ value relies upon theoretical corrections to the adopted line profile shape when measuring $W_{\lambda}$ values. Of the 845 quality-assessable lines, {451} lines are found to have literature atomic data in agreement with the $\Delta\overline{\log(gf)}_{grid}$ values, corresponding to {$\sim$53\%} of the quality-assessable lines. This fraction is also representative of the analysis-independent lines, where {210} lines have sufficiently accurate atomic data. The majority of Fe-group species (\ion{Sc}{II}, \ion{Ti}{I-II}, \ion{V}{I}, \ion{Cr}{I}, \ion{Mn}{I}, \ion{Fe}{II}, \ion{Co}{I}, \ion{Ni}{I}) have a reasonable number of lines with accurate atomic data, typically 70-75\% of lines, however the \ion{Fe}{I} lines only have $\sim$38\% of lines with accurate atomic data.
 
Table~\ref{appendixtable} provides the detailed break down of the findings for the 1091 investigated lines. In an effort to help ascertain the quality of the $\log(gf)$ value of a given spectral line, interested readers can find interactive online plots of all 1091 investigated spectral lines, for all seven benchmark spectra, as hyperlinks in Table~\ref{appendixtable} of the Appendix. The hyperlinks display all cross-matched atomic data for the line in question, including additional information such as level configurations and terms, in addition to interactive plots of the observed line profiles and synthetic line profiles using all $\log(gf)$ values derived here and found in the literature, and their respective equivalent widths per $\log(gf)$ value, per star.

To better visualise the quality assessment process, Figs.~\ref{qa1},~\ref{qa2}, and~\ref{qa3} of the Appendix show the observed lines profiles of three transitions in the seven benchmark spectra, including synthetic line profiles calculated using the $\overline{\log(gf)}_{grid}$ and available literature values. The three transitions show examples of the three quality-assessment outcomes: an example for which all available literature is within the errors of the $\overline{\log(gf)}_{grid}$ value (Fig.~\ref{qa1}); an example for which only some of the available literature values are within the errors of the $\overline{\log(gf)}_{grid}$ value (Fig.~\ref{qa2}); and an example for which none of the available literature values are within the errors of the derived $\overline{\log(gf)}_{grid}$ value (Fig.~\ref{qa3}). It is clear that the $\overline{\log(gf)}_{grid}$ values reproduce the several line profiles extremely well, and that in some cases the available literature data are insufficient to reproduce the observed line profiles.

It is worth mentioning important limitations of this work. The astrophysical $\log(gf)$ values derived in this work were determined using the solar abundances of \cite{solar2007}, and any updates to these elemental abundances would require a correction to the $\log(gf)$ values of lines belonging to updated species. Fortunately, the majority of the lines investigated here would only require directly inverse corrections, that is, if an abundance is revised by $\Delta$[X/H]~=~0.01~dex, then all lines of element X will require a simple update of $\Delta\log(gf)$~=~-0.01~dex. This means that the astrophysical $\log(gf)$ values should remain useful, especially in the context of differential abundance analysis, for years to come. 

Blended lines, even accurately known ones, have been avoided here due to the complexity of simultaneous quality assessment of multiple components. Hyperfine and isotopically split lines have not received any special treatment throughout this analysis. In the case of isotopic lines we find a clear under-representation of s-process elements in the investigated lines. This is because the multiple components are often represented as individual entries in a line list, and so the Section~3.1 selection method treats the components as individual lines, leading to their rejection from analysis due to their high $\Omega_{core}$ values. The hyperfine splitting of lines has not been addressed here, as while hyperfine splitting information is becoming more abundant in the literature, it was not included in the investigated atomic databases at the time of retrieval\footnote{We note that in the time since our atomic data retrievals, VALD3 has expanded their database so that they can now provide HFS data for a large number of atomic species \citep{vald3hfs}.}. Unlike the s-process elements, lines that require hyperfine splitting may be present in the quality assessment results, though inaccuracies in their derived $\log(gf)$ values depend heavily on the observed profile shape. Isotopic and hyperfine lines may benefit from an additional dedicated investigation, where the methods described in this paper are expanded to account for the multiple components of such transitions.

\section{Summary}

In this work we present a homogeneous quality assessment of the literature atomic data available for unblended spectral lines, in the wavelength range 4200-6800~\AA, present in seven benchmark stars that span the effective temperature range 5000-6000~K. The work presented here was performed for multiple benchmark stars to help make the analysis more robust against any erroneous stellar parameter derivations, and also to investigate whether spectral lines are subject to different degrees of blending that would otherwise be unnoticed when analysing a single star. Apart from $T_{\mathrm{eff}}$, the stars are chosen to have similar stellar parameters to minimise any systematic modelling differences between objects. The objects were also chosen to minimise the impact of potential non-LTE effects which could systematically affect the derived $\log(gf)$ values. No clear correlations were found between the derived $\log(gf)$ values and the stellar parameters. The high-quality benchmark spectra were obtained using the Mercator-HERMES spectrograph, each with a resolution of R$\sim$85000 and S/N$\approx$1000. The spectra are available online\footnote{at \href{http://brass.sdf.org}{brass.sdf.org}.} in both normalised and
unnormalised format.

Of the 82337 atomic transitions found in the literature for this wavelength range, 1091 were found to be theoretically deep, unblended, and to have observable line profiles that can be fitted accurately with a single Gaussian profile. Given the high S/N and high resolution of our benchmark spectra, not all of our investigated lines can be used in other spectroscopic works, however the quality assessment results and $\log(gf)$ values themselves are valid at all spectral resolutions and S/Ns. In addition, this work provides quality information for many lines previously unused in spectroscopic analysis.

Astrophysical $\log(gf)$ values were derived for each of the 1091 transitions using two commonly employed methods: equivalent width and spectral synthesis fitting methods. Care was taken to address, minimise, and remove as many potential systematic errors as possible from the derived $\log(gf)$ values, and to produce representative error bars. Agreement between the two derived $\log(gf)$ values was used as a criterion to select a subset of 845 well-behaved lines which could be reliably quality-assessed. An additional constraint, mandating that the $\log(gf)$ values of the two methods are within 0.04~dex of each other, was imposed to produce a further subset of 408 spectral lines which are free of systematic differences caused by the adopted analysis methods.

A key conclusion of this work is that the treatment of blending lines is vital when aiming to reduce systematic errors and uncertainty from quantitative spectroscopy. Careful comparison between the curve of growth and detailed modelling methods revealed that even small background lines, contributing as little as 5\% of the total equivalent width of a spectral feature, can lead to systematic offsets of $\sim$0.10~dex or more between the two methods. 

The available atomic data for these 845 spectral lines were compared against the $\overline{\log(gf)}_{grid}$ values to determine whether the literature values were in agreement with the presented work, as discussed in Section~5.3. It was found that {$\sim$53\%} of the quality-assessed lines have at least one literature $\log(gf)$ value in agreement with our work. For \ion{Fe}{I} this value is $\sim$38\% and for the remaining Fe-group elements the value is around 70-75\% on average. 

We provide recommendations, where possible, as to which literature atomic data will produce sufficiently accurate spectroscopic results.Where the literature is insufficient, such as in some cases where given values are over 0.5~dex from our values, we suggest the use of astrophysical $\log(gf)$ values derived in this work. While the quality assessment presented here is limited to the atomic data compiled in Paper~1, the astrophysical $\log(gf)$ values and errors can easily be used to perform quality assessments of other literature $\log(gf)$ values, allowing astronomers and atomic physicists to evaluate the quality of newly produced atomic data.  All results of the presented investigation are available in the Appendix, as well as in digital format via the CDS and at \href{http://brass.sdf.org}{brass.sdf.org}.

\begin{acknowledgements}
       We thank the atomic data producers and providers for their invaluable work towards improving the accuracy of stellar spectroscopy and the ease at which vast quantities of data can be retrieved. We also thank the anonymous referee for helping to substantially improve the paper. The research for the present results has been subsidised by the Belgian Federal Science policy Office under contract No. BR/143/A2/BRASS. T.M and M.V.d.S are supported by a grant from the Fondation ULB. This work has made use of the VALD database, operated at Uppsala University, the Institute of Astronomy RAS in Moscow, and the University of Vienna. This work is based on observations made with the Mercator Telescope, operated on the island of La Palma by the Flemish Community, at the Spanish Observatorio del Roque de los Muchachos of the Instituto de Astrofísica de Canarias. This work is also based on observations obtained with the HERMES spectrograph, which is supported by the Research Foundation - Flanders (FWO), Belgium, the Research Council of KU Leuven, Belgium, the Fonds National de la Recherche Scientifique (F.R.S.-FNRS), Belgium, the Royal Observatory of Belgium, the Observatoire de Genève, Switzerland and the Thüringer Landessternwarte Tautenburg, Germany. 
\end{acknowledgements}

\bibliographystyle{aa}% style aa.bst
\bibliography{allreferences} % your references Yourfile.bib

\begin{appendix} 

\section{Automatic normalisation of the observed benchmark spectra}

The observed benchmark spectra are continuum flux normalised using a semi-automatic spectral-template normalisation procedure. This procedure searches for a series of wavelength points over sufficiently continuous flux regions close to the stellar continuum level in the theoretical spectra between 4000~\AA\ and 6800~\AA. The selected wavelength regions are then used as continuum anchor points for a polynomial fit to normalise the observed spectrum. We opt to use an automatic template normalisation procedure to remove the `human-factor' from the spectrum normalisation, leading to repeatable and consistent global continuum flux normalisations. The final results are scrutinised to ensure they behave as expected. There are no obvious issues in the continuum placement of the automatically normalised spectra (see Section 4.3.4 for discussion on the $\log(gf)$ uncertainties due to normalisation).

\section{Line profile comparisons with the curve of growth and iterative modelling methods}

\begin{figure*}[h]
    \centering
        \includegraphics[width=18cm,trim={0 1cm 0 1cm},clip]{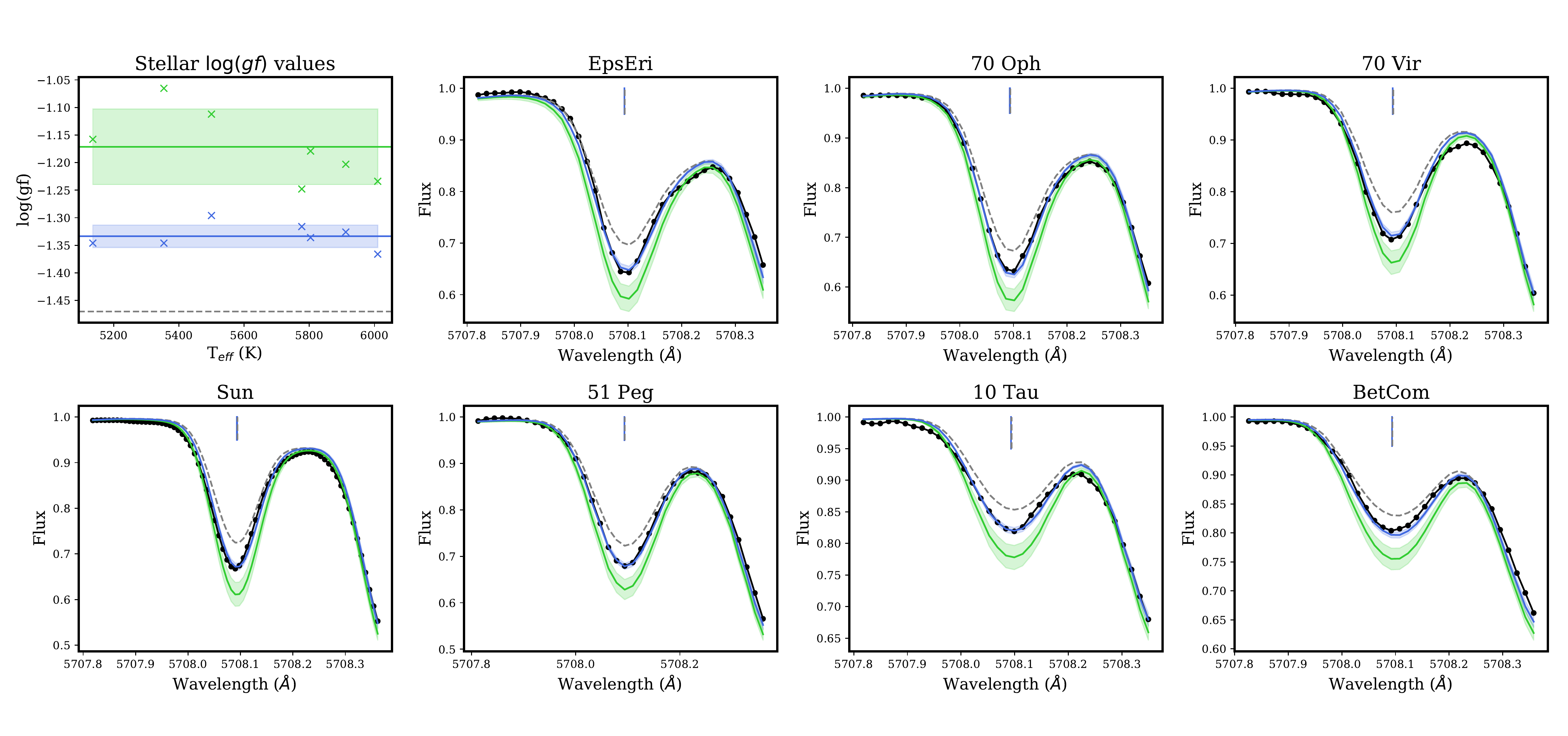}
\caption{ $\log(gf)$ determinations for the $\lambda$5708.1 \ion{Fe}{I} line, which is deemed unsuitable for quality assessment due to the disagreement between the $\overline{\log(gf)}_{cog}$ and $\overline{\log(gf)}_{grid}$ values. Observed line profiles are shown in black dots, line profiles calculated using the $\log(gf)_{input}$ values are shown in dashed grey, line profiles calculated using the $\overline{\log(gf)}_{cog}$ values are shown in green, line profiles calculated using the $\overline{\log(gf)}_{grid}$ values are shown in blue, and errors are shown as green and blue shaded areas, respectively. Vertical blue and dashed grey lines denote the $\overline{\lambda}_{grid}$ and $\lambda_{input}$ values respectively. The disagreement in this case is likely caused by the presence of a nearby spectral line affecting the single Gaussian fit of the curve of growth method. }
\label{invest_example}
\end{figure*}

\begin{figure*}[h]
    \centering
        \includegraphics[width=18cm,trim={0 1cm 0 1cm},clip]{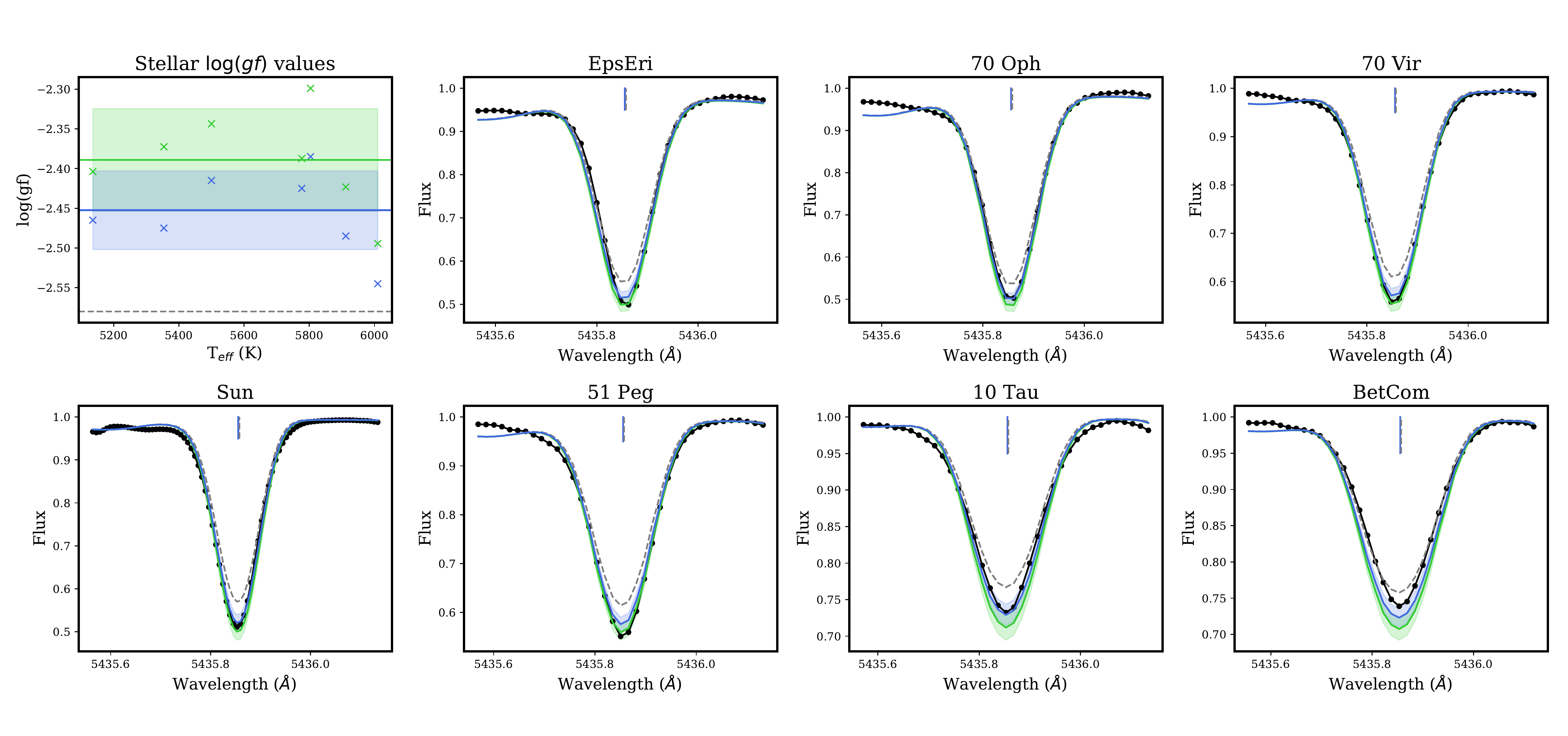}
\caption{ $\log(gf)$ determinations for the $\lambda$5435.8 \ion{Ni}{I} line which is deemed to be quality-assessable. Observed line profiles are shown in black dots, line profiles calculated using the $\log(gf)_{input}$ values are shown in dashed grey, line profiles calculated using the $\overline{\log(gf)}_{cog}$ values are shown in green, line profiles calculated using the $\overline{\log(gf)}_{grid}$ values are shown in blue, and errors are shown as green and blue shaded areas, respectively. Vertical blue and dashed grey lines denote the $\overline{\lambda}_{grid}$ and $\lambda_{input}$ values, respectively. }
\label{qa_example}
\end{figure*}

\begin{figure*}[h]
    \centering
        \includegraphics[width=18cm,trim={0 1cm 0 1cm},clip]{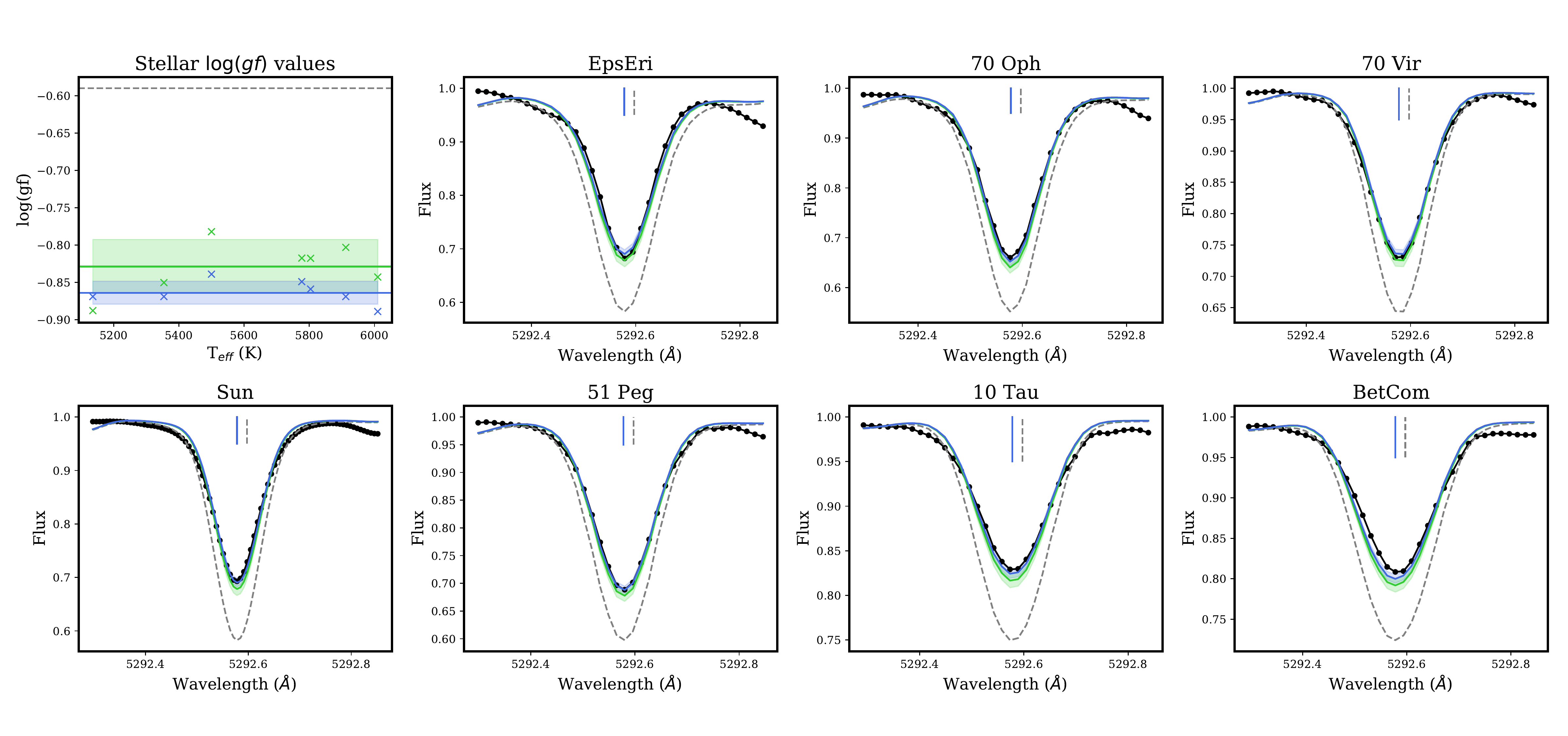}
\caption{ $\log(gf)$ determinations for the $\lambda$5292.6 \ion{Fe}{I} line deemed suitable for quality assessment and deemed to be robust against adopted analysis method. Observed line profiles are shown in black dots, line profiles calculated using the $\log(gf)_{input}$ values are shown in dashed grey, line profiles calculated using the $\overline{\log(gf)}_{cog}$ values are shown in green, line profiles calculated using the $\overline{\log(gf)}_{grid}$ values are shown in blue, and errors are shown as green and blue shaded areas, respectively. Vertical blue and dashed grey lines denote the $\overline{\lambda}_{grid}$ and $\lambda_{input}$ values, respectively. This line also shows a measurable $\Delta\lambda$ correction of $\sim$0.02~\AA\ to the input wavelength. }
\label{robust_example}
\end{figure*}

Figures.~\ref{invest_example},~\ref{qa_example}, and~\ref{robust_example} show the synthesised line profiles of the $\overline{\log(gf)}_{grid}$ and $\overline{\log(gf)}_{cog}$ values, including line profile error bars, alongside the observed line profiles in the seven observed benchmark stars. Figure~\ref{invest_example} shows an example where the $\overline{\log(gf)}_{grid}$ and $\overline{\log(gf)}_{cog}$ values do not agree within error of each other, and thus the spectral line is not suitable for quality assessment. Figure~\ref{qa_example} shows an example where the $\overline{\log(gf)}_{grid}$ and $\overline{\log(gf)}_{cog}$ values do agree within error of each other, making the line suitable for quality assessment. Figure~\ref{qa_example} shows a quality assessable line where the $\overline{\log(gf)}_{grid}$ and $\overline{\log(gf)}_{cog}$ values agree within 0.04~dex of each other, indicating the line is free of systematic issues due to the treatment of line blending by the two methods.

%\newpage

\section{Line profile comparisons with the cross-matched literature data}

\begin{figure*}[h]
    \centering
        \includegraphics[width=18cm,trim={0 1cm 0 1cm},clip]{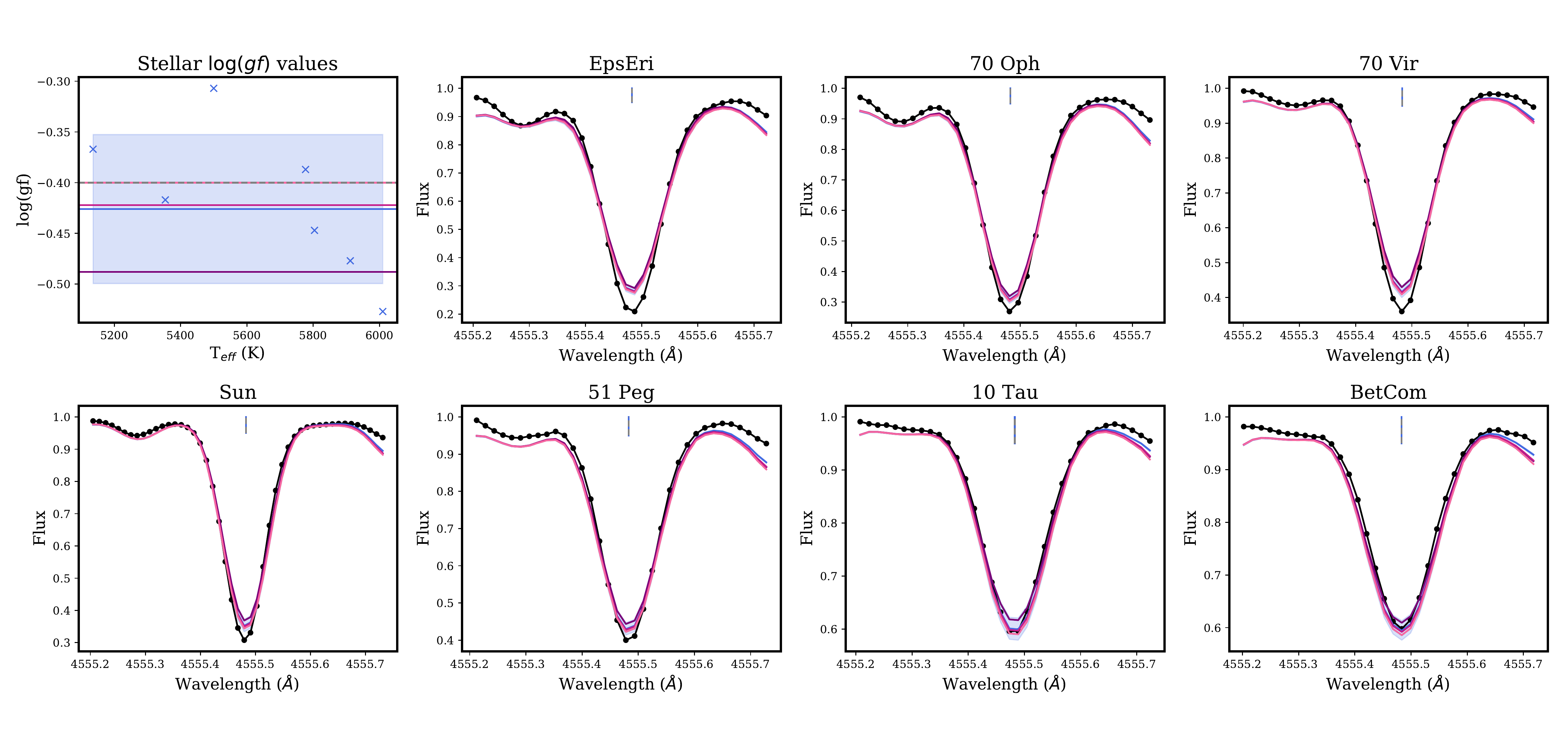}
\caption{Quality assessment for the $\lambda$4555.5 \ion{Ti}{I} line. The style follows that of Figs.~\ref{invest_example}-\ref{robust_example}, minus the $\overline{\log(gf)}_{cog}$ line profiles, and with several additional line profiles corresponding to the available literature $\log(gf)$ values of the transition. In this case all literature $\log(gf)$ values are within error of the $\overline{\log(gf)}_{grid}$ value and can be recommended for use. }
\label{qa1}
\end{figure*}

\begin{figure*}[h]
    \centering
        \includegraphics[width=18cm,trim={0 1cm 0 1cm},clip]{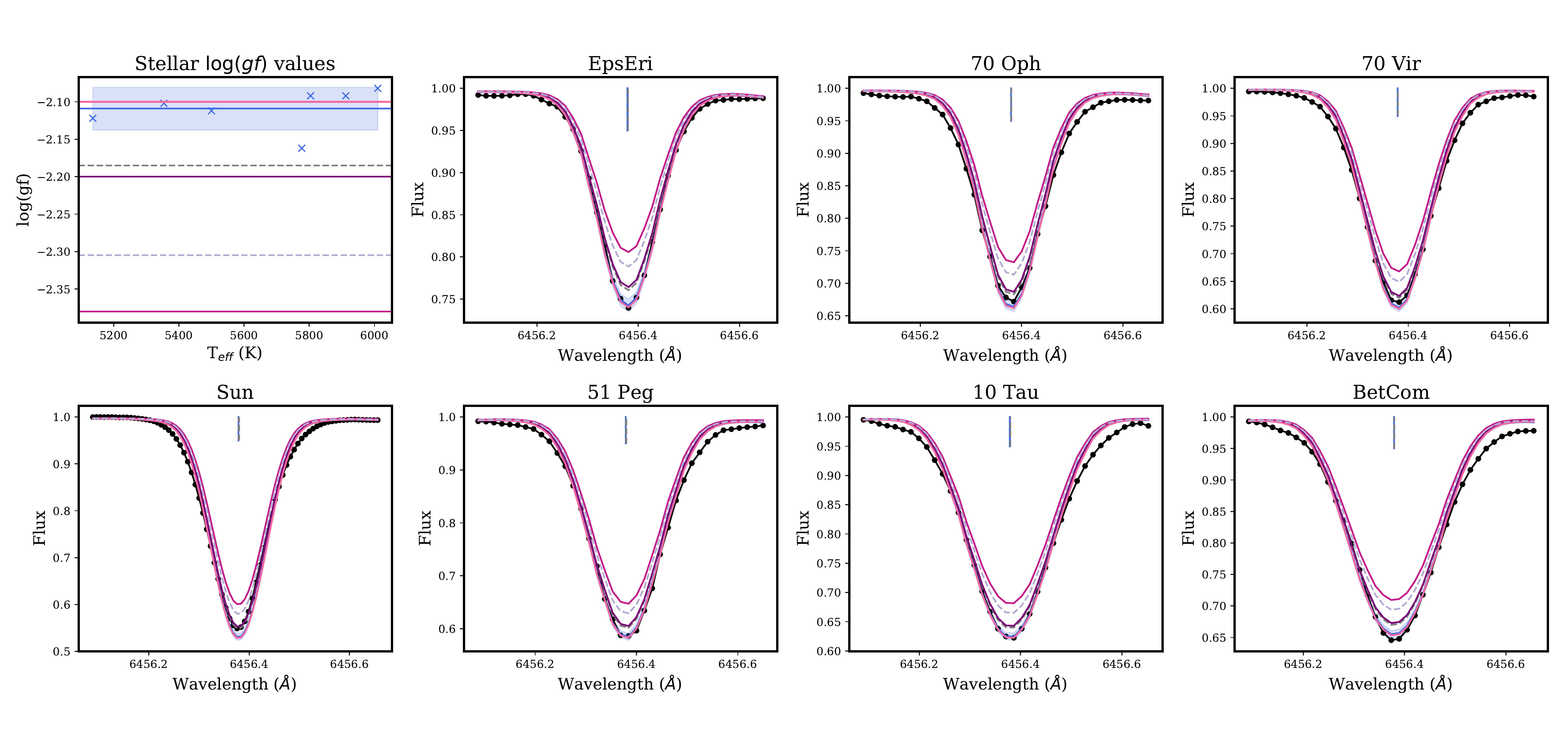}
\caption{Quality assessment for the $\lambda$6456.4 \ion{Fe}{II} line. The style follows that of Fig.~\ref{invest_example}-\ref{robust_example}, minus the $\overline{\log(gf)}_{cog}$ line profiles, and with several additional line profiles corresponding to the available literature $\log(gf)$ values of the transition.  In this case only one literature $\log(gf)$ value is within error of the $\overline{\log(gf)}_{grid}$ value, and thus only this value is recommended for use.  }
\label{qa2}
\end{figure*}

\begin{figure*}[h]
    \centering
        \includegraphics[width=18cm,trim={0 1cm 0 1cm},clip]{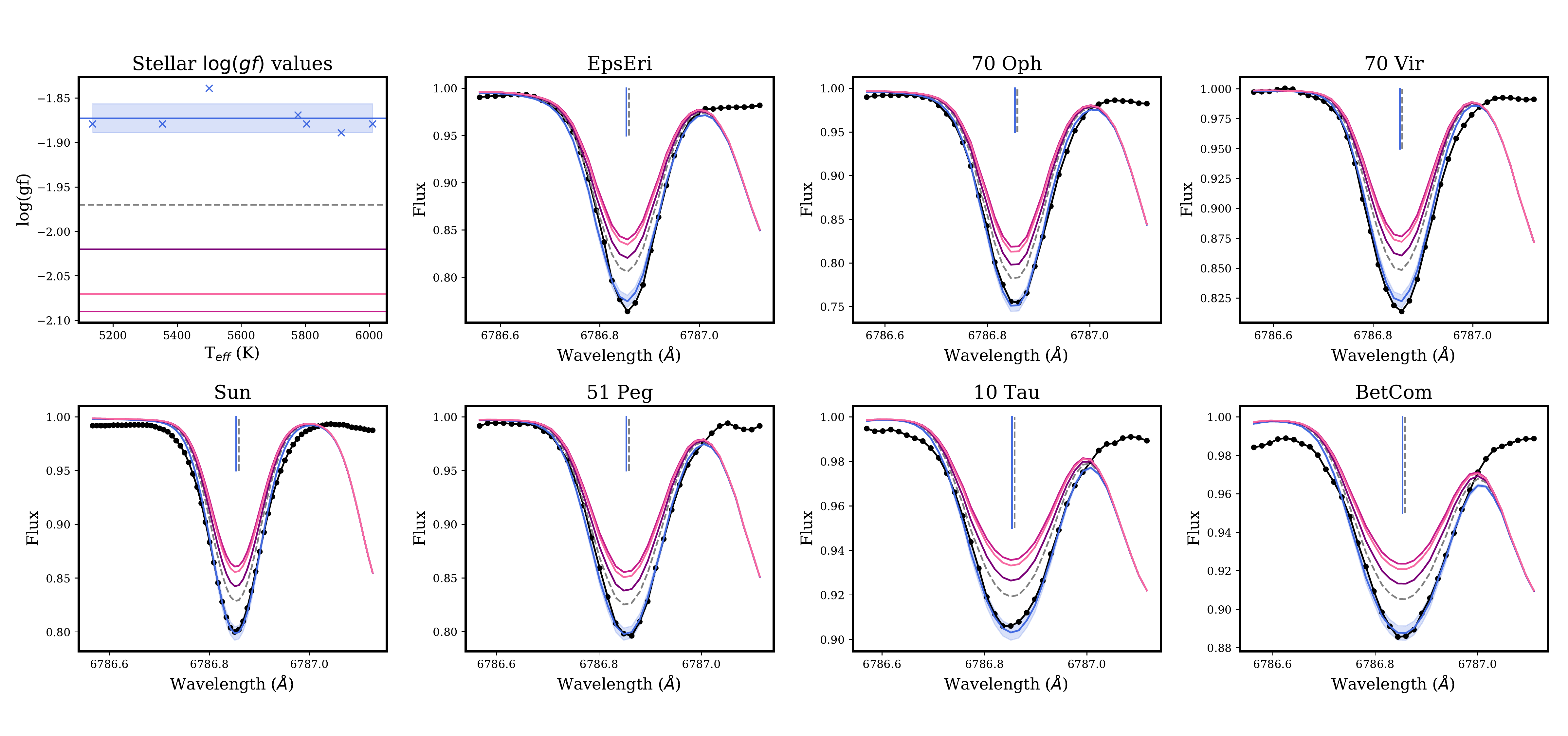}
\caption{Quality assessment for the $\lambda$6786.8 \ion{Fe}{I} line.  The style follows that of Figs.~\ref{invest_example}-\ref{robust_example}, minus the $\overline{\log(gf)}_{cog}$ line profiles, and with several additional line profiles corresponding to the available literature $\log(gf)$ values of the transition. In this case none of the literature $\log(gf)$ values can accurately reproduce the observed profile, with $\Delta\log(gf)$ values ranging between 0.1-0.25~dex, thus we would propose our $\overline{\log(gf)}_{grid}$ as a more accurate alternative. }
\label{qa3}
\end{figure*}

Figures~\ref{qa1},~\ref{qa2},~\ref{qa3} show the synthesised line profiles of the $\overline{\log(gf)}_{grid}$ values, including line profile error bars, alongside the observed line profiles in the seven observed benchmark stars. In addition the figures show synthetic line profiles corresponding to the available literature $\log(gf)$ values of the given transition. All three examples show a analysis-independent line. Figure~\ref{qa1} shows an example where all literature $\log(gf)$ values agree within error of the $\overline{\log(gf)}_{grid}$ value. Figure~\ref{qa2} shows an example where only one of the literature $\log(gf)$ values agrees within error of the $\overline{\log(gf)}_{grid}$ value. Finally, Fig.~\ref{qa3} shows an example where none of the literature $\log(gf)$ values agree within error of the $\overline{\log(gf)}_{grid}$ value.

\section{Tables of individual $\log(gf)_{cog}$ and $\log(gf)_{grid}$ values per benchmark star for the 1091 investigated lines}
The individual $\log(gf)_{cog}$, $\log(gf)_{grid}$, and $\sigma_{grid}$ values per benchmark star are available for each of the 1091 investigated lines in digital format via the CDS and at \href{http://brass.sdf.org}{brass.sdf.org}. Table~\ref{loggfperstar} shows an example of the available individual $\log(gf)_{cog}$ values per benchmark star. Table~\ref{loggfperstar2} shows the same but for the individual $\log(gf)_{grid}$ values, which have the additional $\sigma_{grid}$ values corresponding to the 68.3\% confidence intervals discussed in Section 4.2. 
\begin{table*}
\centering
\caption{An excerpt of the individual $\log(gf)_{cog}$ values derived per benchmark star for each of the 1091 investigated lines. The full table is available in digital format via the CDS and at \href{http://brass.sdf.org}{brass.sdf.org}.}
\begin{tabular}{cccccccc}
            \hline
            \hline
            \noalign{\smallskip}
            Line number & \multicolumn{7}{c}{$\log(gf)_{cog}$}\\
             & $\epsilon$~Eri & 70~Oph~A & 70~Vir & Sun & 51~Peg & 10~Tau & $\beta$~Com \\
                        
            \noalign{\smallskip}
            \hline
            \noalign{\smallskip}
                         1 & -1.72 & -1.66 & -1.41 & -1.47 & -1.54 & -1.46 & -1.49  \\
                         2 & -1.22 & -1.19 & -0.96 & -0.99 & -1.06 & -0.91 & -0.96  \\
                         3 & -0.76 & -0.71 & -0.50 & -0.50 & -0.60 & -0.41 & -0.51  \\
            \noalign{\smallskip}
                        ...&...&...&...&...&...&...&...\\
            \noalign{\smallskip}
            \hline            
\end{tabular}
\label{loggfperstar}
\end{table*}

\begin{table*}
\centering
\caption{An excerpt of the individual $\log(gf)_{grid}$ values, and corresponding $\sigma_{grid}$ values, derived per benchmark star for each of the 1091 investigated lines. The full table is available in digital format via the CDS and at \href{http://brass.sdf.org}{brass.sdf.org}.}
\begin{tabular}{cccccccc}
            \hline
            \hline
            \noalign{\smallskip}
            Line number & \multicolumn{7}{c}{$\log(gf)_{grid}$}\\
             \# & $\epsilon$~Eri & 70~Oph~A & 70~Vir & Sun & 51~Peg & 10~Tau & $\beta$~Com \\
                        
            \noalign{\smallskip}
            \hline
            \noalign{\smallskip}
                         1 & -1.77$_{\pm0.30}$ & -1.72$_{\pm0.30}$ & -1.44$_{\pm0.16}$ & -1.44$_{\pm0.14}$ & -1.56$_{\pm0.23}$ & -1.53$_{\pm0.17}$ & -1.55$_{\pm0.22}$  \\
                         2 & -1.40$_{\pm0.22}$ & -1.26$_{\pm0.22}$ & -0.93$_{\pm0.14}$ & -0.93$_{\pm0.12}$ & -1.00$_{\pm0.14}$ & -0.93$_{\pm0.10}$ & -0.97$_{\pm0.11}$  \\
                         3 & -1.06$_{\pm0.13}$ & -1.00$_{\pm0.16}$ & -0.71$_{\pm0.17}$ & -0.59$_{\pm0.14}$ & -0.73$_{\pm0.17}$ & -0.60$_{\pm0.16}$ & -0.68$_{\pm0.17}$  \\
            \noalign{\smallskip}
                        ...&...&...&...&...&...&...&...\\
            \noalign{\smallskip}
            \hline            
\end{tabular}
\label{loggfperstar2}
\end{table*}

\section{Tables of $W_{\lambda}$ values per benchmark star for the 1091 investigated lines}

The measured $W_{\lambda}$ values, $\Delta W_{\lambda}$ corrections (as discussed in Section 4.3.5), and calculated $W_{\lambda}$ values for the $\overline{\log(gf)}_{grid}$ values, $\overline{\log(gf)}_{cog}$ values, and available literature $\log(gf)$ values are available, for each of the benchmark stars, in digital format via the CDS and at \href{http://brass.sdf.org}{brass.sdf.org}. Table~\ref{equivalentwidthexample} shows an example of the $W_{\lambda}$ values associated with the benchmark star $\epsilon$~Eri for the 1091 investigated spectral lines. Similar tables are available for each of the benchmark stars.

\begin{table*}
\centering
\caption{An excerpt of the individual various $W_{\lambda}$ values available for $\epsilon$~Eri for each of the 1091 investigated lines. The table provides the observed $W_{\lambda}$ values measured using a single Gaussian fit, $\Delta W_{\lambda~corr}$ values used to correct the measured values Gaussian profile assumption (as described in Section 4.3.5), as well as calculated $W_{lambda}$ values for the $\overline{\log(gf)}_{grid}$ values, $\overline{\log(gf)}_{cog}$ values, and available literature $\log(gf)$ values: $^{a)}$ Paper~1 input list~ $^{b)}$ NIST~ $^{c)}$ SpectroWeb~ $^{d)}$ VALD3~ $^{e)}$ Spectr-W$^3$~ $^{f)}$ CHIANTI~ $^{g)}$ TIPbase~ $^{h)}$ TOPbase. Similar tables are available for the remaining benchmark stars. The full tables are available in digital format via the CDS and at \href{http://brass.sdf.org}{brass.sdf.org}.}
\begin{tabular}{ccccccccccccc}
            \hline
            \hline
            \noalign{\smallskip}
            Line number & $W_{\lambda~meas}$ (m\AA) & $\Delta W_{\lambda~corr}$ (m\AA) & \multicolumn{10}{c}{$W_{\lambda~calc}$ (m\AA)}\\
             \#&  &  & grid & cog & a) &  b) &  c) &  d) &  e) &  f) &  g) &  h) \\
                        
            \noalign{\smallskip}
            \hline
            \noalign{\smallskip}
                        1 & 92.8 & +5.3 & 113.5 & 112.1 & 118.4 & - & 118.4 & - &- & - & 60.6 & - \\
                        2 & 119.2 & +13.8 & 166.6 & 158.9 & 117.9 & - & - & - & - & - & - & - \\
                        3 & 175.8 & +27.5 & 200.7 & 248.4 & 272.1 & - & 263.9 & - & 270.4 & - & - & - \\
            \noalign{\smallskip}
                        ...&...&...&...&...&...&...&...&...&...&...&...&...\\
            \noalign{\smallskip}
            \hline            
\end{tabular}
\label{equivalentwidthexample}
\end{table*}

\section{Table of quality assessment information for the 1091 investigated lines}

Table~\ref{appendixtable} presents both the line selection and quality assessment results for the full 1091 spectral lines investigated in this work. The table is available in digital format via the CDS and at \href{http://brass.sdf.org}{brass.sdf.org}.

\longtab[1]{
\begin{landscape}
\begin{longtable}{ccccccccccccccccc}
\caption{Table of results for the 1091 investigated lines in FGK stars, belonging to 28 different ionic species. Lines are listed in elemental and ionic order (from lightest to heaviest, beginning with the neutral species of each element). Each transition shows: $\quad^{1)}$ the corresponding line number, \#, of the transition used to identify the 1091 investigated lines in the various digital tables belonging to this work $\quad^{2)}$ the input wavelength $\quad^{3)}$ the derived $\overline{\lambda}_{grid}$ including 1$\sigma$ error $\quad^{4)}$ the lower energy E$_{low}$ of the transition $\quad^{5)}$ the $\overline{\log(gf)}_{cog}$ and $\overline{\log(gf)}_{grid}$ values derived in this work including 1$\sigma$ errors $\quad^{6)}$ whether a transition is considered quality-assessable (QA) and independent of adopted analysis methods (analysis-independent, or AI) $\quad^{7)}$ the available literature $\log(gf)$ values for a transition taken from Paper~1 including references, where the following columns represent the database of origin for the given literature value: $^{a)}$ Paper~1 input list $^{b)}$ NIST $^{c)}$ SpectroWeb $^{d)}$ VALD3 $^{e)}$ Spectr-W$^3$ $^{f)}$ CHIANTI $^{g)}$ TIPbase $^{h)}$ TOPbase $\quad^{8)}$ which literature references, if any, reproduce the $\overline{\log(gf)}_{grid}$ values within the $\sigma_{grid}$ errors and can thus be recommended for use. If a line cannot be quality-assessed then the 'recommended literature' column will display an 'X', and if the line can be quality-assessed but has no recommended literature values it displays a '-'. Each line number, \#, is hyperlinked to a corresponding web-page where interested readers can find interactive line profile plots of all the $\log(gf)$ values in each benchmark star (in a similar manner to Fig.~\ref{invest_example} and Fig.~\ref{qa1}). Additional atomic information, such as configurations, terms, J-values, and equivalent widths, can be found via these links, or via \href{http://brass.sdf.org}{brass.sdf.org}.
\label{appendixtable}}\\
\hline\hline

\tiny \# & \multicolumn{2}{c}{ Wavelength (\AA)~~~~~~~~~}  &\tiny $E_{low}$ (eV) & \multicolumn{2}{c}{$\log(gf)_{~this~work}$} & \multicolumn{2}{c}{Flags} & \multicolumn{8}{c}{Literature $\log(gf)$ values} & Recommended \\
&\tiny Input &\tiny $\Delta\lambda_{~grid} $ & &\tiny cog &\tiny grid &\tiny QA? &\tiny AI? &\tiny a) &\tiny b) &\tiny c) &\tiny d) &\tiny e) &\tiny f) &\tiny g) &\tiny h) & literature \\
\hline
\noalign{\smallskip}
\endfirsthead
\caption{$^{(QA)}$ Quality-assessable~ $^{(AI)}$ Analysis-independent line~ $^{a)}$ Paper~1 input list~ $^{b)}$ NIST~ $^{c)}$ SpectroWeb~ $^{d)}$ VALD3~ $^{e)}$ Spectr-W$^3$~ $^{f)}$ CHIANTI~ $^{g)}$ TIPbase~ $^{h)}$ TOPbase}\\
\hline\hline
\tiny \# & \multicolumn{2}{c}{ Wavelength (\AA)~~~~~~~~~}  &\tiny $E_{low}$ (eV) & \multicolumn{2}{c}{$\log(gf)_{~this~work}$} & \multicolumn{2}{c}{Flags} & \multicolumn{8}{c}{Literature $\log(gf)$ values} & Recommended \\
&\tiny Input &\tiny $\Delta\lambda_{~grid} $ & &\tiny cog &\tiny grid &\tiny QA? &\tiny AI? &\tiny a) &\tiny b) &\tiny c) &\tiny d) &\tiny e) &\tiny f) &\tiny g) &\tiny h) &\ literature \\
\hline
\noalign{\smallskip}
\endhead
\hline
\endfoot
\noalign{\smallskip}

\multicolumn{17}{c}{\ion{C}{I}}\\
{\tiny \href{http://brass.sdf.org/G_type_htmls/gra409/index.html}{409} } & {\tiny 5052.144} & {\tiny~0.002$_{~\pm0.006}$} & {\tiny7.685} & {\tiny-1.12$_{~\pm0.11}$ } & {\tiny-1.29$_{~\pm0.12}$ } & \ckmk & {\tiny X} & {\tiny-1.30$^{~1}$ } & {\tiny-1.30$^{~1}$ } & {\tiny-1.30$^{~2}$ } & {\tiny-1.30$^{~3}$ } & { \tiny-1.49$^{~4}$ } & - & - & {\tiny-1.45$^{~5}$ } & { \tiny1, 2, 3}\\
\noalign{\smallskip}
\noalign{\smallskip}
\multicolumn{17}{c}{\ion{Na}{I}}\\
{\tiny \href{http://brass.sdf.org/G_type_htmls/gra358/index.html}{358} } & {\tiny 4982.814} & {\tiny~0.004$_{~\pm0.002}$} & {\tiny2.104} & {\tiny-0.76$_{~\pm0.08}$ } & {\tiny-0.86$_{~\pm0.11}$ } & \ckmk & {\tiny X} & {\tiny-0.92$^{~6}$ } & - & {\tiny-0.96$^{~2}$ } & - & - & - & - & - & {\tiny2, 6}\\
{\tiny \href{http://brass.sdf.org/G_type_htmls/gra901/index.html}{901} } & {\tiny 6154.225} & {\tiny-0.006$_{~\pm0.002}$} & {\tiny2.102} & {\tiny-1.42$_{~\pm0.06}$ } & {\tiny-1.45$_{~\pm0.04}$ } & \ckmk & \ckmk & {\tiny-1.55$^{~6}$ } & {\tiny-1.55$^{~7}$ } & {\tiny-1.56$^{~2}$ } & {\tiny-1.55$^{~3}$ } & - & - & - & {\tiny-1.55$^{~5}$ } & -\\
{\tiny \href{http://brass.sdf.org/G_type_htmls/gra905/index.html}{905} } & {\tiny 6160.747} & {\tiny-0.004$_{~\pm0.002}$} & {\tiny2.104} & {\tiny-1.12$_{~\pm0.07}$ } & {\tiny-1.12$_{~\pm0.08}$ } & \ckmk & \ckmk & {\tiny-1.25$^{~6}$ } & {\tiny-1.25$^{~7}$ } & {\tiny-1.26$^{~2}$ } & {\tiny-1.25$^{~3}$ } & - & - & - & {\tiny-1.25$^{~5}$ } & -\\
\noalign{\smallskip}
\noalign{\smallskip}
\multicolumn{17}{c}{\ion{Mg}{I}}\\
{\tiny \href{http://brass.sdf.org/G_type_htmls/gra125/index.html}{125} } & {\tiny 4571.096} & {\tiny-0.008$_{~\pm0.003}$} & {\tiny0.000} & {\tiny-5.46$_{~\pm0.09}$ } & {\tiny-5.57$_{~\pm0.09}$ } & \ckmk & {\tiny X} & {\tiny-5.62$^{~3}$ } & {\tiny-5.62$^{~8}$ } & {\tiny-5.69$^{~2}$ } & {\tiny-5.62$^{~3}$ } & - & - & - & - & {\tiny3, 8}\\
\end{longtable}

\tablebib{

 $^1$~\citet{tableref1};
 $^2$~\citet{spectroweb};
 $^3$~\citet{ralchenko2010};
 $^4$~\citet{wiese1980};
 $^5$~calculated in non-relativistic LS-coupling by \citet{brass1} using \citet{topbase} data;
 $^6$~\citet{tableref6};
 $^7$~\citet{tableref7};
 $^8$~\citet{tableref8};
 $^9$~\citet{tableref9};
 $^{10}$~\citet{tableref10};
 $^{11}$~\citet{tableref11};
 $^{12}$~\citet{tableref12};
 $^{13}$~\citet{wiese1969};
 $^{14}$~\citet{kuruczdata};
 $^{15}$~\citet{tableref15};
 $^{16}$~\citet{garz1973} rescaled using \citet{obrian1991};
 $^{17}$~\citet{garz1973};
 $^{18}$~\citet{tableref18};
 $^{19}$~\citet{tableref19};
 $^{20}$~\citet{chianti};
 $^{21}$~\citet{tableref21};
 $^{22}$~\citet{tableref22};
 $^{23}$~\citet{tableref23};
 $^{24}$~\citet{tableref24};
 $^{25}$~\citet{smith1981};
 $^{26}$~\citet{tableref26};
 $^{27}$~\citet{nicholls1964};
 $^{28}$~\citet{tableref28};
 $^{29}$~\citet{lawler1989};
 $^{30}$~\citet{lawler2013};
 $^{31}$~\citet{blackwell1982ati};
 $^{32}$~\citet{tableref32};
 $^{33}$~\citet{martin1988};
 $^{34}$~\citet{tableref34};
 $^{35}$~\citet{blackwell1986ti};
 $^{36}$~\citet{tableref36};
 $^{37}$~\citet{kostyk1982ti};
 $^{38}$~\citet{blackwell1983ti};
 $^{39}$~\citet{blackwell1986ti} rescaled using \citet{grevesse1989};
 $^{40}$~\citet{nitz1998};
 $^{41}$~\citet{blackwell1983ti} rescaled using \citet{grevesse1989};
 $^{42}$~\citet{tableref42};
 $^{43}$~\citet{tableref43};
 $^{44}$~\citet{tableref44};
 $^{45}$~\citet{wood2013};
 $^{46}$~\citet{pickering2002};
 $^{47}$~\citet{roberts1973};
 $^{48}$~\citet{tableref48};
 $^{49}$~\citet{kostyk1983};
 $^{50}$~\citet{tableref50};
 $^{51}$~\citet{tableref51};
 $^{52}$~\citet{tableref52};
 $^{53}$~\citet{lawler2014};
 $^{54}$~\citet{tableref54};
 $^{55}$~\citet{whaling1985};
 $^{56}$~Bridges (priv. comm. with NIST, 1976);
 $^{57}$~\citet{sobeck2007};
 $^{58}$~\citet{tozzi1985};
 $^{59}$~\citet{tableref59};
 $^{60}$~\citet{kostyk1981};
 $^{61}$~\citet{tableref61};
 $^{62}$~\citet{tableref62};
 $^{63}$~\citet{tableref63};
 $^{64}$~\citet{raassen1998jphb};
 $^{65}$~\citet{booth1984};
 $^{66}$~\citet{tableref66};
 $^{67}$~\citet{fmw1988};
 $^{68}$~ calculated in non-relativistic LS-coupling by \citet{brass1} using \citet{tipbase} data;
 $^{69}$~\citet{obrianplus1991};
 $^{70}$~\citet{blackwell1979};
 $^{71}$~re-normalised values of \citet{mrw1974};
 $^{72}$~\citet{ruffoni2014};
 $^{73}$~\citet{denhartog2014};
 $^{74}$~\citet{blackwell1982fe1};
 $^{75}$~\citet{tableref75};
 $^{76}$~\citet{obrianplus1991} rescaled using \citet{ruffoni2014};
 $^{77}$~\citet{bardkock1994};
 $^{78}$~\citet{tableref78};
 $^{79}$~\citet{tableref79};
 $^{80}$~\citet{tableref80};
 $^{81}$~\citet{bardkock1994} rescaled using \citet{obrianplus1991};
 $^{82}$~\citet{bardkockkock1991} rescaled using \citet{denhartog2014};
 $^{83}$~\citet{blackwell1982fe1} rescaled using \citet{obrianplus1991};
 $^{84}$~\citet{tableref85} rescaled using \citet{obrianplus1991};
 $^{85}$~\citet{tableref85};
 $^{86}$~\citet{bardkockkock1991};
 $^{87}$~\citet{bardkock1994} rescaled using \citet{denhartog2014};
 $^{88}$~\citet{bardkockkock1991} rescaled using \citet{obrianplus1991};
 $^{89}$~\citet{blackwell1980a};
 $^{90}$~\citet{raassen1998};
 $^{91}$~\citet{tableref91};
 $^{92}$~\citet{tableref92};
 $^{93}$~\citet{tableref93};
 $^{94}$~\citet{tableref94};
 $^{95}$~\citet{tableref95};
 $^{96}$~\citet{tableref96};
 $^{97}$~\citet{kostyk1982ni};
 $^{98}$~\citet{wickliffe1997};
 $^{99}$~\citet{wood2014};
 $^{100}$~\citet{doerr1985};
 $^{101}$~\citet{tableref101};
 $^{102}$~re-normalised values of \citet{tableref101};
 $^{103}$~\citet{tableref103};
 $^{104}$~Komarovskiy \& Shabanova (priv. comm with Spectr-W$^3$, 1992);
 $^{105}$~\citet{tableref105};
 $^{106}$~\citet{biemont2011};
 $^{107}$~\citet{hannaford1982};
 $^{108}$~\citet{tableref108};
 $^{109}$~\citet{tableref109};
 $^{110}$~\citet{tableref110};
 $^{111}$~\citet{corliss1962};
 $^{112}$~\citet{tableref112};
 $^{113}$~\citet{tableref113}

}
\end{landscape}
}% End longtab

\end{appendix}

\end{document}